%% file: IntModVerRTMP.tex
\documentclass[a4paper]{article}
\usepackage{times}

\usepackage[pdftex]{graphicx}
\usepackage{rotating}
\DeclareGraphicsExtensions{.ps,.eps,.pdf}
\usepackage{amsmath,amssymb,amsthm}
\usepackage{dsfont}

\usepackage{triotex}

\newcommand{\frf}[1]{(\ref{#1})}
\newcommand{\fsrf}[2]{(\ref{#1}--\ref{#2})}

\newcommand{\becomesMTL}[1]{\MTLoperator{\triangle}{}{}{#1}}
\newcommand{\becomesOMTL}[1]{\MTLoperator{\blacktriangle}{}{}{#1}}
\newcommand{\becomesLMTL}[1]{\becomesOMTL{#1}}

\newcommand{\triggerMTL}[1]{\MTLoperator{\wr\!\wr}{}{}{#1}}
\newcommand{\triggerLMTL}[1]{\MTLoperator{\wr}{}{}{#1}}
\newcommand{\triggerOMTL}[1] {\MTLoperator{\wr\!\!\wr\!\!\wr}{}{}{#1}}
\newcommand{\triggerNOMTL}[1] {\MTLoperator{\bar{\wr\!\!\wr\!\!\wr}}{}{}{#1}}
\newcommand{\triggerNLMTL}[1]{\MTLoperator{\bar{\wr}}{}{}{#1}}
\newcommand{\triggerNMTL}[1]{\MTLoperator{\bar{\wr\!\wr}}{}{}{#1}}

\newcommand{\pre}[1]{\bullet#1}
\newcommand{\post}[1]{#1\bullet}

\newcommand{\tstart}{t_{\mathrm{start}}}

\newcommand{\tait}[1]{\mathrm{#1}}
\newcommand{\idle}{\tait{idle}}
\newcommand{\try}{\tait{try}}

\newcommand{\dataret}{\tait{data\_retrieved}}

\newcommand{\pp}{\mathsf{p}}

\newcommand{\system}{\phi^{\mathsf{sys}}}
\newcommand{\prop}{\phi^{\mathsf{prop}}}

\newcommand{\zot}{$\integers$ot}

\newcommand{\overap}[1]{\mathrm{O}_{\delta}\left({#1}\right)}
\newcommand{\overapempty}{\mathrm{O}_{\delta}}

\newcommand{\underap}[1]{\mathrm{\Omega}_{\delta}\left({#1}\right)}
\newcommand{\underapempty}{\mathrm{\Omega}_{\delta}}

\newcommand{\logictrue}{\top}
\newcommand{\logicfalse}{\bot}

\newcommand{\Jcal}{\mathcal{J}}
\newcommand{\Pcal}{\mathcal{P}}
\newcommand{\Dcal}{\mathcal{D}}
\newcommand{\Ncal}{\mathcal{N}}
\newcommand{\Bchi}{\mathcal{B}_{\chi}}

\newcommand{\modelstime}[1]{\models_{#1}}

\newcommand{\mdrz}{\modelstime{\reals_{\geq 0}}}

\newcommand{\mdt}{\modelstime{\timedomain}}
\newcommand{\behav}{\mathcal{B}}

\newcommand{\MU}[1]{\mu_{#1}}
\newcommand{\TAU}[1]{\tau_{#1}}
\newcommand{\EPSILON}[1]{\epsilon_{#1}}
\newcommand{\ALPHA}[1]{\alpha(#1)}
\newcommand{\BETA}[1]{\beta(#1)}

\theoremstyle{plain}
\newtheorem{theorem}{Theorem}
\newtheorem{proposition}[theorem]{Proposition}

\theoremstyle{definition}

\title{Integrated Modeling and Verification of\\ Real-Time Systems through Multiple Paradigms}

\author{Marcello M. Bersani, Carlo A. Furia, Matteo Pradella, and Matteo Rossi}

\date{7 June 2009}

\begin{document}

\maketitle

\begin{abstract}
Complex systems typically have many different parts and facets, with different characteristics.
In a multi-paradigm approach to modeling, formalisms with different natures are used in combination to describe complementary parts and aspects of the system.
This can have a beneficial impact on the modeling activity, as different paradigms can be better suited to describe different aspects of the system.
While each paradigm provides a different view on the many facets of the system, it is of paramount importance that a coherent comprehensive model emerges from the combination of the various partial descriptions.
In this paper we present a technique to model different aspects of the same system with different formalisms, while keeping the various models tightly integrated with one another.
In addition, our approach leverages the flexibility provided by a bounded satisfiability checker to encode the verification problem of the integrated model in the propositional satisfiability (SAT) problem; this allows users to carry out formal verification activities both on the whole model and on parts thereof.
The effectiveness of the approach is illustrated through the example of a monitoring system.

\end{abstract}

\input{Introduction}

\input{Background}

\input{TPN}

\input{CaseStudy}

\input{Conclusion}


\bibliographystyle{abbrv}
\bibliography{IntModVerRTMP-bib}

\end{document}

%% file: Introduction.tex
\section{Introduction}\label{sec:intro}

Modeling paradigms come in many different flavors: graphical or textual; executable or not; formal, informal, or semi-formal; more or less abstract; with different levels of expressiveness, naturalness, conciseness, etc.
Notations for the design of real-time systems, in addition, include a notion of time, whose characteristics add a further element of differentiation \cite{HM96}.

A common broad categorization of modeling notations distinguishes between \emph{operational} and \emph{descriptive} paradigms \cite{FMMR09}.
Operational notations --- such as Statecharts, finite state automata, or Petri nets --- represent systems through the notions of \emph{state} and \emph{transition} (or event); system behavior consists in evolutions from state to state, triggered by event occurrences.
On the other hand, descriptive paradigms --- such as temporal logics, descriptive logics, or algebraic formalisms --- model systems by declaring their fundamental \emph{properties}.

The distinction between operational and descriptive models is, like with most classifications, neither rigid nor sharp.
None\-the\-less, it is often useful in practice to guide the developer in the choice of notation based on what is being modeled and what are the ultimate goals (and requirements) of the modeling endeavor.
In fact, operational and descriptive notations have different --- and often complementary --- strengths and weaknesses.
Operational models, for instance, are often easier to understand by experts of domains other than computer science (mechanical engineers, control engineers, etc.), which makes them a good design vehicle in the development of complex systems involving components of many different natures.
Also, once an operational model has been built, it is typically straightforward to execute, simulate, animate, or test it.
On the other hand, descriptive notations are the most natural choice when writing partial models of systems, because one can build the description \emph{incrementally} by listing the (partial) known properties one at a time.
For similar reasons, descriptive models are often excellent languages to document the \emph{requirements} of a system: the requirements elicitation process is usually an incremental trial-and-error activity, and thus it benefits greatly from notations which allow cumulative development.

When modeling timed systems, in addition, the choice of the time domain is a crucial one, and it can significantly impact on the features of the model \cite{FMMR09}.
For example, a dense time model is typically needed to represent true asynchrony.
Discrete time, instead, is usually more amenable to automated verification, and is at the basis of a number of quite mature techniques and tools that can be deployed in practice to verify systems.

In this paper we present a technique to model different aspects of the same system with different formalisms, while keeping the various models tightly integrated with one another.
In this approach, modelers can pick their preferred modeling technique and modeling paradigm (e.g., operational or descriptive, continuous or discrete) depending on the particular facet or component of the system to be described.
Integration of the separate snippets in a unique model is made possible by providing a common formal semantics to the different formalisms involved.
Finally, our approach leverages the flexibility provided by a bounded satisfiability checker to encode the verification problem of the integrated model in the propositional satisfiability (SAT) problem; this allows users to carry out formal verification activities both on the whole model and on parts thereof.

The technique presented in this paper hinges on Metric Temporal Logic (MTL) to provide a common semantic foundation to the integrated formalisms, and on the results presented in \cite{FR06} to integrate continuous- and discrete-time MTL fragments into a unique formal description.
Operational formalisms can then be introduced in the framework by providing suitable MTL formalizations, which can then be discretized as well according to the same technique.
While this idea is straightforward in principle, putting it into practice is challenging for several basic reasons.
First, in order to have full discrete-time decidability we have to limit ourselves to \emph{propositional} MTL \cite{AH93}; its relatively limited expressive power makes it arduous to formalize completely the behavior of operational models (some technical facts, briefly described in Section \ref{sec:background}, justify this intuition).
Second, even if we used a more expressive first-order temporal-logic language, formalizing the semantics of ``graphical'' operational formalisms is usually tricky as several semantic subtleties that are ``implicit'' in the original model must be properly understood and resolved when translating them into a logic language.
See for instance extensive discussions of such subtleties in \cite{FMM94} for timed Petri nets and in \cite{vdB94} for Statecharts.
Third, not any MTL axiomatization is amenable to the discretization techniques of \cite{FPR08-FM08}, as syntactically different MTL descriptions yielding the same underlying semantics provide discretizations of wildly different ``qualities''.
Indeed, experience showed that the most ``natural'' axiomatizations of operational formalisms require substantial rewriting in order to work reasonably well under the discretization framework.
Crafting suitable MTL descriptions has proved demanding, delicate, and crucially dependent on the features of the operational formalism at hand.
In this respect, our previous work \cite{FPR08-ICFEM08} focused on a variant of Timed Automata (TA) --- a typical ``synchronous'' operational formalism.
The formalization of intrinsically asynchronous components --- such as those that sit at the boundary between the system and its environment --- demands however the availability of a formalism that is both operational and ``asynchronous''.
To this end, the present paper develops an axiomatization of Timed Petri Nets (TPN), an ``asynchronous'' operational formalism, integrates all three formalisms (MTL, TA, and TPN) into a unique framework, and evaluates an implementation of the framework on a monitoring system example.

The paper is structured as follows.
Section \ref{sec:relatedwork} briefly discusses some works that are related to the approach and technique presented in this article.
Section \ref{sec:background} introduces the relevant results on which the modeling and verification approach presented in this paper are based; more precisely, the section introduces MTL, timed automata and their MTL-based semantics, and the discretization technique for continuous-time MTL formulas.
Section \ref{sec:TPN} presents the (continuous-time) MTL semantics of timed Petri nets and uses it to derive a discretized version of timed Petri nets that can be input to verification engines for discrete-time MTL (e.g., \zot).
Section \ref{sec:casestudy} shows how the various formalisms can be used to describe, and then combine together in a unique model, different aspects and parts of the same system; in addition, it reports on some verification tests carried out on the modeled system.
Finally, Section \ref{sec:conclusion} concludes and outlines some future works in this line of research.

\subsection{Related work} \label{sec:relatedwork} 

Combining different modeling paradigms in a single framework for verification purposes is not a novel concept.
In fact, there is a rich literature on du\-al-lan\-guage approaches, which combine an operational formalism and a descriptive formalism into one analysis framework \cite{FMMR09}.
The operational notation is used to describe the system dynamics, whereas the properties to be checked are expressed through the descriptive notation.
Model-checking techniques \cite{CGP00} are a widely-used example of a dual-language approach to formal verification.
Dual-language frameworks, however, usually adopt a rigid stance, in that one formalism is used to describe the system, while another is used for the properties to be verified. 
In this work we propose a flexible framework in which different paradigms can be mixed for different design purposes: system modeling, property specification and also verification.

Modeling using different paradigms is a staple of UML \cite{UML}.
In fact, the UML modeling language is actually a blend of different notations (message sequence charts, Statecharts, OCL formulas, etc.) with different characteristics.
The UML framework provides means to describe the same (software) systems from different, possibly complementary, perspectives.
However, the standard language is devoid of mechanisms to guarantee that an \emph{integrated} global view emerges from the various documents or that, in other words, the union of the different views yields a precise, coherent model.

Some work has been devoted to the (structural) transformation between models to re-use verification techniques for different paradigms and to achieve a unified semantics, similarly to the approach of this paper.
Cassez and Roux \cite{CR06} provide a structural translation of TPN into TA that allows one to piggy-back the efficient model-checking tools for TA.
Our approach is complementary to \cite{CR06} and similar works\footnote{See the related work section of \cite{CR06} for more examples of transformational approaches.} in several ways.
First, our transformations are targeted to a discretization framework: on the one hand, this allows a more lightweight verification process as well as the inclusion of discrete-time components within the global model; on the other hand, discretization introduces incompleteness that might reduce its effectiveness.
Second, we leverage on a descriptive notation (MTL) rather than an operational one.
This allows the seamless integration of operational and descriptive components, whereas the transformation of \cite{CR06} stays within the model-checking paradigm where the system is modeled within the operational domain and the verified properties are modeled with a descriptive notation.
Also, state-of-the-art of tools for model-checking of TA (and formalisms of similar expressive power) do not support full real-time temporal logics (such as TCTL) but only a subset of significantly reduced expressive power.
We claim that the model and properties we consider in the example of Section \ref{sec:casestudy} are rather sophisticated and deep---even after weighting in the inherent limitations of our verification technique.

For the sake of brevity, we omit in this report a description of related works on the discretization of continuous-time models.
The interested reader can refer to \cite{FPR08-FM08} for a discussion of this topic.


%% file: Background.tex
\section{Background}\label{sec:background}

\subsection{Continuous- and discrete-time real-time behaviors}
%
We represent the concept of \emph{trace} (or \emph{run}) of some real-time system through the notion of \emph{behavior}.
Given a time domain $\timedomain$ and a finite set $\Pcal$ of atomic propositions, a behavior $b$ is a mapping $b: \timedomain \rightarrow 2^\Pcal$ which associates with every time instant $t \in \timedomain$ the set $b(t)$ of propositions that hold at $t$.
$\behav_{\timedomain}$ denotes the set of all behaviors over $\timedomain$ (for an implicit fixed set of propositions).
$t \in \timedomain$ is a \emph{transition point} for behavior $b$ iff $t$ is a discontinuity point of the mapping $b$.
Depending on whether $\timedomain$ is a discrete, dense, or continuous set, we call a behavior over $\timedomain$ discrete-, dense-, or continuous-time respectively.
In this report, we assume the natural numbers $\naturals$ as discrete time domain and the nonnegative real numbers $\reals_{\geq 0}$ as continuous (and dense) time domain.

\paragraph{Non-Zeno and non-Berkeley.}
Over continuous-time domains, it is customary to consider only physically meaningful behaviors, namely those respecting the so-called non-Zeno property.
A continuous-time behavior $b$ is non-Zeno if the sequence of transition points of $b$ has no accumulation points.
For a non-Zeno behavior $b$, it is well-defined the notions of values to the left and to the right of any transition point $t > 0$, which we denote as $b^-(t)$ and $b^+(t)$, respectively.
When a proposition $p \in \Pcal$ is such that $p \in b^-(t) \Leftrightarrow p \not\in b^+(t)$ (i.e., $p$ switches its truth value about $t$), we say that $p$ is ``triggered'' at $t$.
In order to ensure reducibility between continuous and discrete time, we consider non-Zeno behaviors with a stronger constraint, called \emph{non-Berkeleyness}.
A continuous-time behavior $b$ is non-Berkeley for some positive constant $\delta \in \reals_{> 0}$ if, for all $t \in \timedomain$, there exists a closed interval $[u, u+\delta]$ of size $\delta$ such that $t \in [u, u+\delta]$ and $b$ is constant throughout $[u, u+\delta]$.
Notice that a non-Berkeley behavior (for any $\delta$) is non-Zeno \emph{a fortiori}. 
The set of all non-Berkeley continuous-time behaviors for $\delta > 0$ is denoted by $\Bchi^\delta \subset \behav_{\reals_{\geq0}}$.
In the following we always assume behaviors to be non-Berkeley, unless explicitly stated otherwise.

\paragraph{Syntax and semantics.}
From a purely semantic point of view, one can consider the model of a (real-time) system simply as a set of behaviors \cite{AH92b,FMMR07-TR2007-22} over some time domain $\timedomain$ and sets of propositions.
In practice, however, systems are modeled through some suitable notation: in this paper we consider a mixture of MTL formulas \cite{Koy90,AH93}, TA \cite{AD94,AFH96}, and TPN \cite{CMS99}.
Given an MTL formula, a TA, or a TPN $\mu$, and a behavior $b$, $b \models \mu$ denotes that $b$ represents a system evolution which satisfies all the constraints imposed by $\mu$.
If $b \models \mu$ for some $b \in \behav_{\timedomain}$, $\mu$ is called $\timedomain$-satisfiable; if $b \models \mu$ for all $b \in \behav_{\timedomain}$, $\mu$ is called $\timedomain$-valid.
Similarly, if $b \models \mu$ for some $b \in \Bchi^\delta$, $\mu$ is called $\chi^\delta$-satisfiable; if $b \models \mu$ for all $b \in \Bchi^\delta$, $\mu$ is called $\chi^\delta$-valid.

\subsection{Descriptive notation: Metric Temporal Logic}
Let $\Pcal$ be a finite (non-empty) set of atomic propositions and $\Jcal$ be the set of all (possibly unbounded) intervals of the time domain $\timedomain$ with rational endpoints.
We abbreviate intervals with pseudo-arithmetic expressions, such as $=d$, $<d$, $\geq d$, for $[d,d]$, $(0,d)$, and $[d, +\infty)$, respectively.

\paragraph{MTL syntax.}
The following grammar defines the syntax of (propositional) MTL, where $I \in \Jcal$ and $\pp \in \Pcal$. 
\begin{equation*}
  \phi  ::=  \pp  \mid  \neg \phi  \mid  \phi_1 \wedge \phi_2  \mid
                   \untilMTL{I}{\phi_1, \phi_2}  \mid \sinceMTL{I}{\phi_1, \phi_2}
\end{equation*}

The basic temporal operators of MTL is the \emph{bounded until} $\untilMTL{I}{\phi_1, \phi_2}$ (and its past counterpart \emph{bounded since} $\sinceMTL{I}{}$) which says that $\phi_1$ holds until $\phi_2$ holds, with the additional constraint that $\phi_2$ must hold within interval $I$.
Throughout the paper we omit the explicit treatment of past operators (i.e., $\sinceMTL{I}{}$ and derived) as it can be trivially derived from that of the corresponding future operators.

\paragraph{MTL semantics.}
MTL semantics is defined over behaviors, parametrically with respect to the choice of the time domain $\timedomain$.
While the semantics of Boolean connectives and In particular, the definition of the \emph{until} operators is as follows:
\\
\begin{tabular}{l c l}
  $b(t) \mdt \untilMTL{I}{\phi_1, \phi_2}$ & \ \ \ iff\ \ \ &
            there exists $d \in I$ such that: $b(t+d) \mdt \phi_2$  \\
  &  &      and, for all $u \in [0, d]$ it is $b(t+u) \mdt \phi_1$  \\
   $b \mdt \phi$  & \ \ \ iff\ \ \ &  for all $t \in \timedomain$: $b(t) \mdt \phi$
\end{tabular}

We remark that a global satisfiability semantics is assumed, i.e., the satisfiability of formulas is implicitly evaluated over \emph{all} time instants in the time domain.
This permits the direct and natural expression of most common real-time specifications (e.g., time-bounded response, time-bounded invariance, etc.) without resorting to nesting of temporal operators.

\paragraph{Granularity.}
For an MTL formula $\phi$, let $\Jcal_{\phi}$ be the set of all non-null, finite interval bounds appearing in $\phi$.
Then, $\Dcal_\phi$ is the set of positive values $\delta$ such that any interval bound in $\Jcal_{\phi}$ is an integer if divided by $\delta$.

\subsubsection{Derived (temporal) operators.}
It is customary to introduce a number of derived (temporal) operators, to be used as shorthands in writing specification formulas.
We assume a number of standard abbreviations such as $\logicfalse, \logictrue, \vee, \Rightarrow, \Leftrightarrow$; when $I = (0, \infty)$, we drop the subscript interval in temporal operators.
All other derived operators used in this paper are listed in Table \ref{tab:mtl-derived} ($\delta \in \reals_{> 0}$ is a parameter used in the discretization techniques, discussed shortly).
In the following we describe briefly and informally the purpose of such derived operators, focusing on future ones (the meaning of the corresponding past operators is easily derivable).

\begin{itemize}

\item For propositions in the set $\{ \gamma(x) \mid x \in X \}$, $\bigodot_{x \in X \subseteq Y} \gamma(x)$ states  that $\gamma(x)$ holds for all $x$ in $X$ and does not hold for all $x$ in the complement set $Y \setminus X$.

\item A few common derived temporal operators such as $\relMTL{I}{}, \diamondMTL{I}{}, \boxMTL{I}{}$ are defined with the usual meaning: $\relMTL{I}{}$ (\emph{release}) is the dual of the \emph{until} operator; $\diamondMTL{I}{\phi}$ means that $\phi$ happens within time interval $I$ in the future; $\boxMTL{I}{\phi}$ means that $\phi$ holds throughout the whole interval $I$ in the future.

\item $\nowonstrMTL{\phi}$ and $\nowonMTL{\phi}$ are useful over continuous time only, and describe $\phi$ holding throughout some unspecified non-empty interval in the strict future; more precisely, if $t$ is the current instant, there exists some $t' > t$ such that $\phi$ holds over $\langle t, t')$, where the interval is left-open for $\nowonstrMTL{}$ and left-closed for $\nowonMTL{}$.

\item $\becomesMTL{}$ and $\becomesOMTL{}$ describe different types of \emph{transitions}. Namely, $\becomesMTL{\phi_1, \phi_2}$ describes a switch from $\phi_1$ to $\phi_2$, irrespective of which value holds at the current instant, whereas $\becomesOMTL{\phi_1, \phi_2}$ describes a switch from $\phi_1$ to $\phi_2$ such that $\phi_1$ holds at the current instant and $\phi_2$ will hold in the immediate future. Note that if $\becomesMTL{\phi_1, \phi_2}$ holds at some instant $t$, $\becomesOMTL{\phi_1, \phi_2}$ holds over $(t-\delta, t)$.

\item $\becomesMTL{\phi}, \becomesOMTL{\phi}$ are shorthands for transitions of a single item; correspondingly the $\triggerLMTL{}, \triggerLMTL{}, \triggerOMTL{}$ ``trigger'' operators are introduced: $\triggerLMTL{\phi}$ denotes a transition of $\phi$ from false to true or \emph{vice versa}, whereas $\triggerMTL{\phi}$ describes a similar transition where the value of $\phi$ at the current instant is unspecified.
  $\triggerOMTL{\phi' \leadsto \phi}$ describes a more complex transition of $\phi$, one which is ``triggered'' by the auxiliary proposition $\phi'$.

\item It is also convenient to introduce the ``dual'' operators
$\triggerNLMTL{}, \triggerNOMTL{}$ which describe ``non-transitions'' of
their argument. For instance, $\triggerNLMTL{\phi}$ says that the truth
value of $\phi$ (whatever it is) does not change from the current instant to the immediate future.

\item Finally, $\Alw{\phi}$ expresses the invariance of $\phi$. Since $b \mdt \Alw{\phi}$ iff $b \mdt \phi$, for any behavior $b$, $\Alw{\phi}$ can be expressed without nesting if $\phi$ is flat, through the global satisfiability semantics introduced beforehand.
\end{itemize}

\begin{table}[!htb]
\begin{scriptsize}
\begin{center}
  \begin{tabular}{|c @{$\quad \equiv \quad$} c|}
	 \hline
    \textsc{Operator}        & \textsc{Definition}  \\
    \hline
    $\bigodot_{x \in X \subseteq Y} \gamma(x)$   & $\bigwedge_{x \in X} \gamma(x) \:\wedge\: \bigwedge_{y \in Y \setminus X} \neg \gamma(y)$ \\
    \hline
	 $\relMTL{I}{\phi_1, \phi_2}$    &   $\neg \untilMTL{I}{\neg \phi_1, \neg \phi_2}$  \\
	 $\redMTL{I}{\phi_1, \phi_2}$    &   $\neg \sinceMTL{I}{\neg \phi_1, \neg \phi_2}$  \\
	 $\diamondMTL{I}{\phi}$    &   $\untilMTL{I}{\logictrue, \phi}$  \\
	 $\diamondPMTL{I}{\phi}$    &   $\sinceMTL{I}{\logictrue, \phi}$ \\
	 $\boxMTL{I}{\phi}$    &   $\relMTL{I}{\logicfalse, \phi}$ \\
	 $\boxPMTL{I}{\phi}$    &   $\redMTL{I}{\logicfalse, \phi}$ \\
    \hline
    $\nowonstrMTL{\phi}$     &    $\untilMTL{(0, +\infty)}{\phi, \logictrue} \vee (\neg \phi \wedge \relMTL{(0, +\infty)}{\phi, \logicfalse})$  \\
    $\uptonowstrMTL{\phi}$     &    $\sinceMTL{(0, +\infty)}{\phi, \logictrue} \vee (\neg \phi \wedge \redMTL{(0, +\infty)}{\phi, \logicfalse})$ \\
	 $\nowonMTL{\phi}$  &  $\phi \wedge \nowonstrMTL{\phi}$ \\
	 $\uptonowMTL{\phi}$  &  $\phi \wedge \uptonowstrMTL{\phi}$ \\
    \hline
	 $\becomesMTL{\phi_1, \phi_2}$ &   $\begin{cases}
                                  \uptonowstrMTL{\phi_1} \wedge \left( \phi_2 \vee \nowonstrMTL{\phi_2} \right)   &
                                   \text{if } \timedomain = \reals_{\geq 0} \\
							  \diamondPMTL{=1}{\phi_1} \wedge \diamondMTL{[0, 1]}{\phi_2}    &
                                   \text{if } \timedomain = \naturals
                                         \end{cases}$ \\
	 $\becomesOMTL{\phi_1, \phi_2}$  &  $\begin{cases}
                       \phi_1 \wedge \diamondMTL{=\delta}{\phi_2} &
                                   \text{if } \timedomain = \reals_{\geq 0} \\
							  \phi_1 \wedge \diamondMTL{=1}{\phi_2}    &
                                   \text{if } \timedomain = \naturals
                                          \end{cases}$ \\
    \hline
    $\becomesMTL{\phi}$  &  $\becomesMTL{\neg \phi, \phi}$ \\
    $\becomesOMTL{\phi}$  &  $\becomesOMTL{\neg \phi, \phi}$ \\
    $\triggerLMTL{\phi}$  &  $\becomesLMTL{\phi} \vee \becomesLMTL{\neg \phi}$ \\
    $\triggerMTL{\phi}$  &  $\becomesMTL{\phi} \vee \becomesMTL{\neg \phi}$ \\
    $\triggerOMTL{\phi' \leadsto \phi}$  &  $\begin{cases}
                                 \uptonowstrMTL{\neg \phi} \wedge \boxMTL{=\delta}{\phi' \Rightarrow \phi} \:\vee\: \uptonowstrMTL{\phi} \wedge \boxMTL{=\delta}{\phi' \Rightarrow \neg \phi}   &      \text{if } \timedomain = \reals_{\geq 0} \\
                                 \boxPMTL{[0,1]}{\neg \phi} \wedge \boxMTL{[0,2]}{\phi' \Rightarrow \phi} \vee \boxPMTL{[0,1]}{\phi} \wedge \boxMTL{[0,2]}{\phi' \Rightarrow \neg \phi}         &       \text{if } \timedomain = \naturals
                         \end{cases}$  \\
    \hline
    $\triggerNMTL{\phi}$ & $\becomesMTL{\phi, \phi}$ \\
    $\triggerNLMTL{\phi}$  &  $\becomesLMTL{\phi, \phi} \vee \becomesLMTL{\neg \phi, \neg\phi}$ \\
    $\triggerNOMTL{\phi' \leadsto \phi}$  &  $\begin{cases}
                                 \uptonowstrMTL{\phi} \wedge \boxMTL{=\delta}{\phi' \Rightarrow \phi} \:\vee\: \uptonowstrMTL{\neg \phi} \wedge \boxMTL{=\delta}{\phi' \Rightarrow \neg \phi}   &      \text{if } \timedomain = \reals_{\geq 0} \\
                                 \boxPMTL{[0,1]}{\phi} \wedge \boxMTL{[0,2]}{\phi' \Rightarrow \phi} \vee \boxPMTL{[0,1]}{\neg\phi} \wedge \boxMTL{[0,2]}{\phi' \Rightarrow \neg \phi}         &       \text{if } \timedomain = \naturals
                         \end{cases}$  \\
    \hline
    $\Alw{\phi}$   &   $\phi \wedge \boxMTL{(0, +\infty)}{\phi} \wedge \boxPMTL{(0, +\infty)}{\phi}$ \\
   \hline
  \end{tabular}
  \caption{MTL derived temporal operators}
  \label{tab:mtl-derived}
\end{center}
\end{scriptsize}
\end{table}

\subsection{Operational notations: Timed Automata and Timed Petri Nets} \label{sec:timedautomata}
For lack of space, we omit a formal presentation of TA, which have been however introduced in the framework in previous work \cite{FPR08-ICFEM08} and focus on MTL and TPN in the following.
Section \ref{sec:casestudy} will however informally illustrate the syntax and semantics of TA on an example, with a level of detail sufficient to understand its role within the framework.

\paragraph{Timed Petri nets syntax.}
A \emph{Timed Petri Net} (TPN) is a tuple $N = \langle P, T, F, M_0, \alpha, \beta \rangle$:
\begin{itemize}
\item $P$ is a finite set of \emph{places};
\item $T$ is a finite set of \emph{transitions};
\item $F \subseteq (P \times T) \cup (T \times P)$ is the \emph{flow relation};
\item $M_0: P \rightarrow \naturals$ is the \emph{initial marking};
\item $\alpha: T \rightarrow \rationals_{\geq 0}$ gives the \emph{earliest firing times} of transitions; and
\item $\beta: T \rightarrow \rationals_{\geq 0} \cup \{\infty\}$ gives the \emph{latest firing times} of transitions.
\end{itemize}

In general, a mapping $M: P \rightarrow \naturals$ is called a \emph{marking} of $N$.
Given $a \in P \cup T$, let $\pre{a} = \{ b \mid b F a \}$  and $\post{a} = \{b \mid a F b \}$ denote the preset and postset of $a$, respectively.
We assume that every node $a \in P \cup T$ has a nonempty preset or a nonempty postset (or both); this is clearly without loss of generality.

\paragraph{Timed Petri nets semantics.}
The semantics of TPN is usually given as sequences of transition firings and place markings; see \cite{CMS99} for formal definitions.
Correspondingly, a TPN is called \emph{$k$-safe} for $k \in \naturals$ iff for every reachable marking $M$ it is $M(p) \leq k$ for all $p \in P$.
A TPN that is $k$-safe for some $k \in \naturals$ is called \emph{bounded}.

In this report we assume \emph{1-safe TPN}.
This allows a simplified description of the semantics, where any marking is completely described by a set $M \subseteq P$ of places such that a place is marked iff it is in $M$.
We remark, however, that extending the presentation to generic \emph{bounded} TPN would be routine.
On the other hand, unbounded TPN would not be discretizable according to the notion of Section \ref{sec:discretization}, hence they would fit only in a different framework.
To further simplify the presentation, we assume non-Berkeley behaviors for some generic $\delta > 0$ in presenting the semantics; correspondingly we do not have to consider zero-time transitions as every enabled transition is enabled for at least $\delta$ time units.

The continuous-time semantics of a 1-safe TPN $N = \langle P, T, F, M_0, \alpha, \beta \rangle$ can be conveniently introduced for behaviors over propositions in $\Pcal = \mu \,\cup\, \epsilon \,\cup\, \tau = \{ \mu(p), \epsilon(p) \mid p \in P\} \cup \{ \tau(t) \mid t \in T\}$ as follows.
Intuitively, at any time $t$ over a behavior $b$, $\mu(p) \in b(t)$ denotes that place $p$ is marked; $\tau(u)$ being triggered at $t$ denotes that transition $u$ fires at $t$; and $\epsilon(p)$ being triggered at $t$ denotes that place $p$ undergoes a ``zero-time unmarking'', as it will be defined shortly.\footnote{The dual ``zero-time markings'' do not occur over non-Berkeley behaviors as a consequence of zero-time transitions not occurring.}
Then, $b$ is a \emph{run} of TPN $N$, and we write $b \mdrz N$, iff the following conditions hold:
\begin{itemize}

\item \emph{Initialization}: $b(0) = \epsilon \cup \tau \cup \bigcup_{p \in M_0} \mu(p) $, and there exists a transition instant $\tstart > 0$\footnote{In the following, we will assume that $\tstart \in [0,2\delta]$ for the discretization parameter $\delta > 0$.} such that: $b(t) = (t)$ for all $0 \leq t \leq \tstart$ and $b^+(\tstart) = \tau \cup \bigcup_{p \in M_0} \mu(p)$.

\item \emph{Marking}: for all instants $u > \tstart$ such that $\mu(p) \not\in b^-(u)$ and $\mu(p) \in b^+(u)$ we say that $p$ becomes marked. Correspondingly, there exists a transition $t \in \pre{p}$ such that: (i) $\tau(t)$ is triggered at $u$, (ii) for no other transition $t' \in \pre{p}$ (other than $t$ itself) $\tau(t')$ is triggered at $u$, and (iii) for no transition $t \in \post{p}$ $\tau(t)$ is triggered at $u$.

\item \emph{Unmarking}: for all instants $u > \tstart$ such that $\mu(p) \in b^-(u)$ and $\mu(p) \not\in b^+(u)$ we say that $p$ becomes unmarked. Correspondingly, there exists a transition $t \in \post{p}$ such that: (i) $\tau(t)$ is triggered at $u$, (ii) for no other transition $t' \in \post{p}$ (other than $t$ itself) $\tau(t')$ is triggered at $u$, and (iii) for no transition $t \in \pre{p}$ $\tau(t)$ is triggered at $u$.

\item \emph{Enabling}: for all instants $u > \tstart$ such that $\tau(t)$ is triggered at $u$, all places $p \in \pre{t}$ must have been marked continuously over $(u-\alpha(t), u)$ without any zero-time unmarkings of the same places occurring.

\item \emph{Bound}: for all instants $u > \tstart$ such that $\tau(t)$ has
not been triggered anywhere over $(u-\beta(t), u)$ and all places
$p\in\pre{t}$ have been marked continuously, one of the following must
occur: (i) all such $p$'s becomes unmarked at $u$, (ii) $\tau(t)$ is triggered at $u$, or (iii) all such $p$'s are still marked ``now on'' and some $p \in \pre{t}$ undergoes a zero-time unmarking (i.e., $\epsilon(p)$ is triggered at $u$).

\item \emph{Effect}: for all instants $u > \tstart$ such that $\tau(t)$ is triggered at $u$, any place $p\in\pre{t}$ becomes unmarked or undergoes a zero-time unmarking, and any place $p\in\post{t}$ becomes marked or undergoes a zero-time unmarking.

\item \emph{Zero-time unmarking}: for all instants $u > \tstart$ such that $\epsilon(p)$ is triggered at $u$ we say that $p$ undergoes a zero-time unmarking.
  Correspondingly, there exist transitions $t_a \in \pre{p}$ and $t_b \in \post{p}$ such that $\tau(t_a)$ is triggered, $\tau(t_b)$ is triggered, and for no other transition $t' \in \pre{p} \cup \post{p}$ (other than $t_a,t_b$) $\tau(t')$ is triggered.
\end{itemize}

\subsection{Discrete-time approximations of continuous-time specifications} \label{sec:discretization}

This section provides an overview of the results in \cite{FPR08-FM08} that will be used as a basis for the technique of this paper.
The technique of \cite{FPR08-FM08} is based on two approximation functions for MTL formulas, called under- and over-approximation. 
The under-approximation function $\underapempty$ maps continuous-time MTL formulas to discrete-time formulas such that the non-validity of the latter implies the non-validity of the former, over behaviors in $\Bchi^\delta$; in other words $\underapempty$ preserves validity from continuous to discrete time.
The over-approximation function $\overapempty$ maps continuous-time MTL formulas to discrete-time MTL formulas such that the validity of the latter implies the validity of the former, over behaviors in $\Bchi^\delta$.
We have the following fundamental verification result, which constitutes the basis of the whole verification framework in the paper.

\begin{proposition}[Approximations \cite{FPR08-FM08}] \label{prop:approximations}
  For any MTL formulas $\phi_1, \phi_2$, and for any $\delta \in \Dcal_{\phi_1, \phi_2}$:
  (1) if $\Alw{{\underap{\phi_1}}} \Rightarrow \Alw{\overap{\phi_2}}$ is $\naturals$-valid,
		    then $\Alw{\phi_1} \Rightarrow \Alw{\phi_2}$ is $\chi^\delta$-valid;
and (2) if $\Alw{\overap{\phi_1}} \Rightarrow \Alw{\underap{\phi_2}}$ is not $\naturals$-valid,
		    then $\Alw{\phi_1} \Rightarrow \Alw{\phi_2}$ is not $\chi^\delta$-valid.
\end{proposition}

Proposition \ref{prop:approximations} suggests the following verification approach for MTL.
Assume first a system modeled as an (arbitrarily complex) MTL formula $\system$; in order to verify if another MTL formula $\prop$ holds for all run of the system we should check the \emph{validity} of the derived MTL formula $\Alw{\system} \Rightarrow \Alw{\prop}$ which postulates that every run of the system also satisfies the property.
Over continuous time, we would build the two discrete-time formulas of Proposition \ref{prop:approximations} and infer the validity of the continuous-time formula from the results of a discrete-time validity checking.
The technique is incomplete as, in particular, when approximation (1) is not valid and approximation (2) is valid nothing can be inferred about the validity of the property in the original system over continuous time.

Consider now another notation $\Ncal$ (e.g., TA or TPN); if we can characterize the con\-tin\-u\-ous-time semantics of any system described with $\Ncal$ by means of a set of MTL formulas, we can reduce the (continuous-time) verification problem for $\Ncal$ to the (con\-tin\-u\-ous-time) verification problem for MTL, and solve the latter as outlined in the previous paragraph.

There are, however, several practical hurdles that make this approach not straightforward to achieve.
First, the application of the over- and under- approximations of \cite{FPR08-FM08} requires MTL formulas written in a particular form and which do not nest temporal operators.
Although in principle every formula can be transformed in the required form (possibly with the addition of a finite number of fresh propositional variables), not any transformation is effective.
That is, it turns out that semantically equivalent continuous-time formulas can yield dramatically different --- in terms of efficacy and completeness --- approximated discrete-time formulas.
The axiomatization of operational formalisms (such as TA and TPN) is all the more extremely tricky and requires different sets of axioms, according to whether they will undergo under- or over- approximation.
However, all different axiomatizations will be shown to be continuous-time
equivalent, hence the intended semantics is captured correctly in all situations.
The application in practice of the MTL verification technique will use the ``best'' set of axioms in every case.


%% file: TPN.tex
\section{Discretizable MTL Axiomatizations of TPN}\label{sec:TPN}
It is not too hard to devise a general, continuous-time axiomatization of the semantics of a non-trivial subclass of TPN.
However, this axiomatization---for reasons that are similar to those discussed in \cite{FPR08-ICFEM08} for the TA axiomatization---yields a poor discretized counterpart when the technique of Section \ref{sec:discretization} is applied.
Then, this section describes three equivalent (for non-Berkeley behaviors) continuous-time axiomatizations of the semantics of TPN (as introduced in Section \ref{sec:timedautomata}): a generic one (Section \ref{sec:generic}), one that works best for discrete-time under-approximation (Section \ref{sec:4underap}), and one that works best for discrete-time over-approximation (Section \ref{sec:4overap}).
Sections \ref{sec:UA} and \ref{sec:OA} produce respectively the corresponding discrete-time formulas that will be used in the verification problem.
Throughout this section, assume a TPN $N = \langle P, T, F, M_0, \alpha, \beta \rangle$ and the set of propositions $\Pcal = \mu \cup \epsilon \cup \tau$ as in the definition of their semantics (Section \ref{sec:timedautomata}).
The axiomatization of TPN presented in this paper imposes that, in every marking, a place can contain at most one token.
As a consequence, it captures all evolutions of any TPN that is 1-safe; however, it is also capable of describing, for a TPN that is not 1-safe (i.e., which has reachable markings such that at least one place contains more than one token) the sequences of markings in which every place has at most one token.
For 1-safe TPN (either by construction or by imposition) any marking $M$ is completely described by the subset of places that are marked in $M$, which simplifies their formalization.
We remark, however, that extending the axiomatization to include generic \emph{bounded} TPN would be routine.

\subsection{Generic axiomatization} \label{sec:generic}

The continuous-time semantics of a 1-safe TPN $N = \langle P, T, F, M_0, \alpha, \beta \rangle$ can be described through the set of propositions $\Pcal = \mu \,\cup\, \epsilon \,\cup\, \tau$, where $ \mu = \{ \MU{p} \mid p \in P\}$, $ \epsilon = \{ \EPSILON{p} \mid p \in P\}$ and $\tau = \{ \TAU{u} \mid u \in T\}$.
Intuitively, at any time $t$ in a behavior $b$, $\MU{p} \in b(t)$ denotes that place $p$ is marked; $\TAU{u}$ being ``triggered'' (see Section \ref{sec:background}) at $t$ denotes that transition $u$ fires at $t$; and $\EPSILON{p}$ being triggered at $t$ denotes that place $p$ undergoes a ``zero-time unmarking'', that is, $p$ is both unmarked and marked at the same instant (hence does not change the number of contained tokens), as it will be defined shortly.\footnote{The dual ``zero-time markings'' (in which a place $p$ is both marked and unmarked at the same instant, and hence remains empty) do not occur over non-Berkeley behaviors since, over these behaviors, transitions cannot fire in the same instant in which they are enabled.}
Then, $b$ is a \emph{run} of TPN $N$, and we write $b \mdrz N$, iff the conditions listed below hold.

\subsubsection{Places}
Marking and unmarking of place $p \in P$ is described by linking transitions of $\MU{p}$ to transitions of $\TAU{u}$ for transitions $u$ in the pre and postset of $p$.
The trigger operator $\triggerMTL{}$ (matching $\becomesMTL{}$) is used for $\TAU{u}$ as the actual truth value of $\TAU{u}$ after the transition is irrelevant as long as a transition occurs.

\paragraph{Marking:}
For all instants $t$ such that $\MU{p}$ becomes true in $t$ we say that $p$ becomes marked. Correspondingly, there exists a transition $u \in \pre{p}$ such that: (i) $\TAU{u}$ is triggered at $t$, (ii) for no other transition $u' \in \pre{p}$ (other than $u$ itself) $\TAU{u'}$ is triggered at $t$, and (iii) for no transition $u'' \in \post{p}$ $\tau(u'')$ is triggered at $t$.
This corresponds to the following axioms. 

\begin{align}
  \label{ax:u2m_i}  
  p \in M_0 &: \;
  \becomesMTL{\MU{p}} \;\Rightarrow\; 
  \left(
  \begin{array}{c}
    \bigvee_{u \in \pre{p}} \left(
    \triggerMTL{\TAU{u}}
    \wedge
    \bigwedge_{u'\neq u \in \pre{p}} \triggerNMTL{\TAU{u'}} \right)   
    \wedge
    \bigwedge_{u \in \post{p}} \triggerNMTL{\TAU{u}}       
    \\ \vee \\ 
    \boxPMTL{(0, \infty)}{\neg\MU{p}}
  \end{array} \right)
\\
 \label{ax:u2m}
  p \notin M_0 &: \;  
  \becomesMTL{\MU{p}} \;\Rightarrow\;
    \bigvee_{u \in \pre{p}} 
    \left(
    \triggerMTL{\TAU{u}}
    \wedge
    \bigwedge_{u'\neq u \in \pre{p}} \triggerNMTL{\TAU{u'}}
    \right) 
     \wedge
    \bigwedge_{u \in \post{p}} \triggerNMTL{\TAU{u}}
\end{align}

\paragraph{Unmarking:}
For all instants $t$ such that $\MU{p}$ becomes false in $t$ we say that $p$ becomes unmarked. Correspondingly, there exists a transition $u \in \post{p}$ such that: (i) $\TAU{u}$ is triggered at $t$, (ii) for no other transition $u' \in \post{p}$ (other than $u$ itself) $\TAU{u'}$ is triggered at $t$, and (iii) for no transition $u'' \in \pre{p}$ $\tau(u'')$ is triggered at $t$.

\begin{equation} \label{ax:m2u}
  \becomesMTL{\neg\MU{p}} \quad\Rightarrow\quad
  \bigvee_{u \in \post{p}} 
  \left(
    \triggerMTL{\TAU{u}}
    \;\wedge\;
    \bigwedge_{u'\neq u \in \post{p}} \triggerNMTL{\TAU{u'}}
  \right) \wedge
  \bigwedge_{u \in \pre{p}} \triggerNMTL{\TAU{u}}
\end{equation}

\subsubsection{Transitions}
The lower and upper bounds on the firing of transition $u$ are specified by necessary and sufficient conditions, respectively, on transitions of proposition $\TAU{u}$.
Earliest and latest firing times are introduced through MTL real-time constraints.
A non-firing transition $u$ stays enabled as long as $\MU{p}$ (for $p$ in $t$'s preset) holds continuously.

\paragraph{Enabling:} 
For all instants $t$ such that $\TAU{u}$ is triggered at $t$, all places $p \in \pre{u}$ must have been marked continuously over $(t-\ALPHA{u}, t)$ without any zero-time unmarkings of the same places occurring.

\begin{equation} \label{ax:enabling}
  \triggerMTL{\TAU{u}} \quad\Rightarrow\quad
  \bigwedge_{p \in \pre{u}}
  \left( 
    \begin{array}{c}
      \uptonowstrMTL{\MU{p} \wedge \EPSILON{p}} \wedge \boxPMTL{(0,\ALPHA{u})}{ \MU{p} \wedge \EPSILON{p}}
      \\ \vee \\
      \uptonowstrMTL{\MU{p} \wedge \neg \EPSILON{p}} \wedge \boxPMTL{(0,\ALPHA{u})}{ \MU{p} \wedge \neg\EPSILON{p}}
    \end{array} 
  \right)
\end{equation}

\paragraph{Bound:}
For all instants $t$ such that $\TAU{u}$ has not been triggered anywhere over $(t-\BETA{u}, t)$ and all places $p\in\pre{u}$ have been marked continuously, one of the following must occur: (i) one of such $p$'s becomes unmarked at $t$, (ii) $\TAU{u}$ is triggered at $t$, or (iii) all such $p$'s are still marked in $b^+(t)$ and some $p \in \pre{u}$ undergoes a zero-time unmarking (i.e., $\EPSILON{p}$ is triggered at $t$).
%
This is formalized by introducing two axioms for each transition $u \in T$.

\begin{scriptsize}
\begin{equation} \label{ax:boundP}
  \boxPMTL{(0,\BETA{u})}{\TAU{u} \wedge \bigwedge_{p \in \pre{u}} \MU{p}}
  \quad\Rightarrow\quad
  \left(
  \begin{array}{c}
    \bigvee_{p \in \pre{u}}(\neg\MU{p} \vee \nowonstrMTL{\neg\MU{p}})
  \\ \vee \\
  \begin{array}{c}
  \bigvee_{p \in \pre{u}}
  \left( \begin{array}{c}
      \boxPMTL{(0,\BETA{u})}{\EPSILON{p}} \ \Rightarrow\ \neg\EPSILON{p} \vee \nowonstrMTL{\neg\EPSILON{p}} 
      \\ \wedge \\
      \boxPMTL{(0,\BETA{u})}{\neg\EPSILON{p}} \ \Rightarrow\ \EPSILON{p} \vee \nowonstrMTL{\EPSILON{p}} 
    \end{array}
  \right) \end{array}
   \\ \vee \\
   \neg \TAU{u} 
   \vee    
   \nowonstrMTL{\neg \TAU{u}} 
 \end{array}
\right)
\end{equation}
\end{scriptsize}

\begin{scriptsize}
\begin{equation} \label{ax:boundN}
  \boxPMTL{(0,\BETA{u})}{\neg \TAU{u} \wedge \bigwedge_{p \in \pre{u}} \MU{p}}
  \ \Rightarrow\ 
  \left(
  \begin{array}{c}
    \bigvee_{p \in \pre{u}}(\neg\MU{p} \vee \nowonstrMTL{\neg\MU{p}})
  \\ \vee \\
  \begin{array}{c}
  \bigvee_{p \in \pre{u}}
  \left( \begin{array}{c}
      \boxPMTL{(0,\BETA{u})}{\EPSILON{p}} \ \Rightarrow\ \neg\EPSILON{p} \vee \nowonstrMTL{\neg\EPSILON{p}} 
      \\ \wedge \\
      \boxPMTL{(0,\BETA{u})}{\neg\EPSILON{p}} \ \Rightarrow\ \EPSILON{p} \vee \nowonstrMTL{\EPSILON{p}} 
    \end{array}
  \right) \end{array} 
   \\ \vee \\
   \TAU{u} 
   \vee 
   \nowonstrMTL{\TAU{u}} 
 \end{array}
\right)
\end{equation}
\end{scriptsize}

Axioms (\ref{ax:boundP}--\ref{ax:boundN}) impose a so-called ``strong time semantics'' to the TPN model \cite{FMM94}. 
This is a departure from the notion of TA formalized in \cite{FPR08-ICFEM08}, for which the axioms impose what is in fact a weak time semantics \cite{FMMR09}. 

\paragraph{Effect:} 
For all instants $t$ such that $\TAU{u}$ is triggered at $t$, every place $p\in\pre{u}$ either becomes unmarked or undergoes a zero-time unmarking, and every place $p\in\post{u}$ either becomes marked or undergoes a zero-time unmarking.

\begin{equation} \label{ax:trans2}
  \triggerMTL{\TAU{u}} \quad\Rightarrow\quad
  \bigwedge_{p \in \pre{u}}\left( \begin{array}{c}
                          \becomesMTL{\neg \MU{p}}  \vee  \triggerMTL{\EPSILON{p}} 
                         \end{array} \right)
  \;\wedge\;
  \bigwedge_{p \in \post{u}}\left( \begin{array}{c}
                          \becomesMTL{\MU{p}}  \vee  \triggerMTL{\EPSILON{p}}
                         \end{array} \right)
\end{equation}


\subsubsection{Zero-time unmarking}
For all instants $t$ such that $\EPSILON{p}$ is triggered at $t$ we say that $p$ undergoes a zero-time unmarking.
Correspondingly, there exist transitions $u_a \in \pre{p}$ and $u_b \in \post{p}$ such that $\TAU{u_a}$ is triggered, $\TAU{u_b}$
is triggered, and for no other transition $u' \in \pre{p} \cup \post{p}$ (other than $u_a,u_b$) $\TAU{u'}$ is triggered.

\begin{equation} \label{ax:unm}
  \triggerMTL{\EPSILON{p}} \quad\Rightarrow\quad
  \bigvee_{\substack{u_a \in \pre{p} \\ u_b \in \post{p}}}
  \left( \begin{array}{c}
      \triggerMTL{\TAU{u_a}}
      \;\wedge\;
      \bigwedge_{u'\neq u_a \in \pre{p}} \triggerNMTL{\TAU{u'}}   
      \\ \wedge \\
      \triggerMTL{\TAU{u_b}}
      \;\wedge\;
      \bigwedge_{u'\neq u_b \in \post{p}} \triggerNMTL{\TAU{u'}}
    \end{array} \right) 
\end{equation}


\subsubsection{Initialization}
$b(0) = \epsilon \cup \tau$, and there exists a transition instant $\tstart > 0$
such that: $b(t) = b(0)$ for all $0 \leq t < \tstart$ and $b^+(\tstart) = \epsilon \cup \tau \cup \bigcup_{p \in M_0} \MU{p}$ (i.e., the places in the initial marking become marked at $\tstart$).
This is captured by the following axiom:
\begin{equation} \label{ax:init}
  \text{at $0$: } 
  \quad \bigwedge_{p \in P}\neg\MU{p}
  \wedge \diamondMTL{[0,2\delta]}{\bigwedge_{p \in M_0} \MU{p}} 
  \ \wedge \ \nowonMTL{\bigwedge_{p \in P} \EPSILON{p} \wedge \bigwedge_{u \in T} \TAU{u}}
\end{equation}

Finally, given a TPN N, the MTL formula $\psi_N$ formalizing N is the conjunction of axioms \fsrf{ax:u2m_i}{ax:init} instantiated for each place and transition of N.

\subsection{Axiomatization for under-approximation} \label{sec:4underap}
As also discussed in \cite{FPR08-ICFEM08}, operator $\becomesMTL{}$ yields very weak under-approximations when used to the left-hand side of implications.
It turns out that the under-approximation of $\becomesMTL{\phi_1,\phi_2}$ is the discrete-time formula $\boxPMTL{[0,1]}{\phi_1} \wedge \phi_2$.
For a proposition $x$, $\becomesMTL{x}$ is then the unsatisfiable formula $\boxPMTL{[0,1]}{\neg x} \wedge x$; correspondingly all implications with such formulas as antecedent are trivially true and do not constrain in any way the discrete-time system.

The approximations can be significantly improved by using the more constraining $\becomesLMTL{}$ in place of $\becomesMTL{}$.
One can check that the under-approximation of $\becomesLMTL{x}$ is $\becomesLMTL{x}$ itself, which describes a discrete-time transition with $\neg x$ holding at the current instant and $x$ holding at the next instant.
Correspondingly, all instances of $\becomesMTL{}$ are changed into instances of $\becomesLMTL{}$ in \fsrf{ax:u2m}{ax:init} yielding \fsrf{ax:u2m4UA}{ax:init4UA}.

\subsubsection{Places}

\begin{gather}
  p \in M_0:
  \becomesLMTL{\MU{p}} \Rightarrow
  \left( 
  \begin{array}{c}
    \bigvee_{u \in \pre{p}} 
    \left(
    \triggerLMTL{\TAU{u}}
     \wedge 
    \bigwedge_{u'\neq u \in \pre{p}} \triggerNLMTL{\TAU{u'}}
    \right) \wedge
    \bigwedge_{u \in \post{p}} \triggerNLMTL{\TAU{u}}
    \\\vee\\ 
    \boxPMTL{[0, \infty)}{\neg\MU{p}}  
    \end{array} \right)
 \label{ax:u2m4UA_i} \\
  p \notin M_0:
  \becomesLMTL{\MU{p}} \quad\Rightarrow\quad   
  \begin{array}{c}
    \bigvee_{u \in \pre{p}} 
    \left(
    \triggerLMTL{\TAU{u}}
    \;\wedge\;
    \bigwedge_{u'\neq u \in \pre{p}} \triggerNLMTL{\TAU{u'}}
    \right) \wedge
    \bigwedge_{u \in \post{p}} \triggerNLMTL{\TAU{u}}
  \end{array} 
 \label{ax:u2m4UA} \\
 \becomesLMTL{\neg\MU{p}} \quad\Rightarrow\quad
  \bigvee_{u \in \post{p}} 
  \left(
    \triggerLMTL{\TAU{u}}
    \;\wedge\;
    \bigwedge_{u'\neq u \in \post{p}} \triggerNLMTL{\TAU{u'}}
  \right) \wedge
  \bigwedge_{u \in \pre{p}} \triggerNLMTL{\TAU{u}}
  \label{ax:m2u4UA}
\end{gather}

\subsubsection{Transitions}

\begin{gather}
  \triggerLMTL{\TAU{u}} \quad\Rightarrow\quad
  \bigwedge_{p \in \pre{u}}
  \left( 
    \begin{array}{c}
      \MU{p} \wedge \EPSILON{p} \wedge \boxPMTL{(0,\ALPHA{u} -\delta)}{ \MU{p} \wedge \EPSILON{p}}
      \\ \vee \\
      \MU{p} \wedge \neg \EPSILON{p} \wedge \boxPMTL{(0,\ALPHA{u} -\delta)}{ \MU{p} \wedge \neg\EPSILON{p}}
    \end{array} 
  \right)
   \label{ax:enabling4UA} \\
\text{Same as }\frf{ax:boundP} \label{ax:boundP4UA} \\
\text{Same as }\frf{ax:boundN} \label{ax:boundN4UA} \\
\triggerLMTL{\TAU{u}} \quad\Rightarrow\quad
\bigwedge_{p \in \pre{u}}\left( \begin{array}{c}
    \becomesLMTL{\neg \MU{p}}  \vee  \triggerLMTL{\EPSILON{p}} 
  \end{array} \right)
\;\wedge\;
\bigwedge_{p \in \post{u}}\left( \begin{array}{c}
    \becomesLMTL{\MU{p}}  \vee  \triggerLMTL{\EPSILON{p}}
  \end{array} \right)
\label{ax:trans24UA}
\end{gather}

\subsubsection{Zero-time unmarking}
\begin{gather}
  \triggerLMTL{\EPSILON{p}} \quad\Rightarrow\quad
  \bigvee_{\substack{u_a \in \pre{p} \\ u_b \in \post{p}}}
  \left( \begin{array}{c}
      \triggerLMTL{\TAU{u_a}}
      \;\wedge\;
      \bigwedge_{u'\neq u_a \in \pre{p}} \triggerNLMTL{\TAU{u'}}   
      \\ \wedge \\
      \triggerLMTL{\TAU{u_b}}
      \;\wedge\;
      \bigwedge_{u'\neq u_b \in \post{p}} \triggerNLMTL{\TAU{u'}}
    \end{array} \right) 
\label{ax:unm4UA}
\end{gather}

\subsubsection{Initialization}
\begin{gather}
  \boxPMTL{[\delta, \infty)}{\perp} 
  \quad \Rightarrow \quad
  \bigwedge_{p \in P}\neg\MU{p}
  \wedge \diamondMTL{[0,2\delta]}{\bigwedge_{p \in M_0} \MU{p}} 
  \ \wedge \ \nowonMTL{\bigwedge_{p \in P} \EPSILON{p} \wedge \bigwedge_{u \in T} \TAU{u}}
\label{ax:init4UA}
\end{gather}

It can be shown that \fsrf{ax:u2m_i}{ax:init} are equivalent to \fsrf{ax:u2m4UA_i}{ax:init4UA} over behaviors that are non-Berkeley for $\delta$.
For instance, consider \frf{ax:u2m} and \frf{ax:u2m4UA}.
In order to show that \frf{ax:u2m} implies \frf{ax:u2m4UA}, let \frf{ax:u2m} and $\becomesLMTL{\MU{p}}$ hold at the current time instant $z$.
$\becomesLMTL{\MU{p}}$ implies that there exists a $z' \in [z,z+\delta]$ where $\MU{p}$ shifts from false to true.
\frf{ax:u2m} evaluated at $z'$ entails (among other things) that $\triggerMTL{\TAU{u}}$ holds at $z'$ for some $t$; that is, $\TAU{u}$ is triggered at $z'$.
Without loss of generality, assume that $\TAU{u}$ is false before $z'$ and is true after it.
The non-Berkeleyness assumption allows us to strengthen this fact, so that $\TAU{u}$ is false at $z$ as well and is true until $z+\delta$, because $z' \in [z, z+\delta]$.
Hence $\triggerLMTL{\TAU{u}}$ holds at $z$.
The rest of the implication is proved similarly.
The proof of the converse implication that \frf{ax:u2m4UA} implies \frf{ax:u2m} also relies on the non-Berkeleyness assumption, which guarantees that there is exactly one transition of $\MU{p}$ over $[z,z+\delta]$ as a consequence of $\becomesLMTL{\MU{p}}$ holding at $z$.
We omit the details of the proof, which are however along the same lines.

\subsection{Under-approximation} \label{sec:UA}
The under-approximations of \fsrf{ax:u2m4UA_i}{ax:init4UA} are reported as formulas \fsrf{ax:u2m-UA_i}{ax:init-UA}.
Notice the lower- and upper-bound relaxations in \fsrf{ax:enabling-UA}{ax:boundN-UA}, in accordance with the notion of under-ap\-prox\-i\-ma\-tion.

\subsubsection{Places}

\begin{gather}
 \text{Syntactically the same as in }\frf{ax:u2m4UA_i}
 \label{ax:u2m-UA_i} \\
  \text{Syntactically the same as in }\frf{ax:u2m4UA}
 \label{ax:u2m-UA} \\
 \text{Syntactically the same as in }\frf{ax:m2u4UA}
 \label{ax:m2u-UA}
\end{gather}

\subsubsection{Transitions}

\begin{scriptsize}
\begin{gather}
  \triggerLMTL{\TAU{u}} \quad\Rightarrow\quad
  \bigwedge_{p \in \pre{u}}
  \left( 
    \begin{array}{c}
      \MU{p} \wedge \EPSILON{p} \wedge \boxPMTL{[1,\ALPHA{u}/\delta -2]}{ \MU{p} \wedge \EPSILON{p}}
      \\ \vee \\
      \MU{p} \wedge \neg \EPSILON{p} \wedge \boxPMTL{[1,\ALPHA{u}/\delta -2]}{ \MU{p} \wedge \neg\EPSILON{p}}
    \end{array} 
  \right)
   \label{ax:enabling-UA} \\
  \boxPMTL{[0,\BETA{u}/\delta]}{\TAU{u} \wedge \bigwedge_{p \in \pre{u}} \MU{p}}
  \quad\Rightarrow\quad
  \left(
  \begin{array}{c}
    \bigvee_{p \in \pre{u}}\diamondMTL{=1}{\neg\MU{p}}
  \\ \vee \\
%
  \bigvee_{p \in \pre{u}}
  \left( \begin{array}{c}
      \boxPMTL{[0,\BETA{u}/\delta]}{\EPSILON{p}} \ \Rightarrow\ \diamondMTL{=1}{\neg\EPSILON{p}} 
      \\ \wedge \\
      \boxPMTL{[0,\BETA{u}/\delta]}{\neg\EPSILON{p}} \ \Rightarrow\ \diamondMTL{=1}{\EPSILON{p}} 
    \end{array}
  \right) 
   \\ \vee \\
   \diamondMTL{=1}{\neg \TAU{u}} 
 \end{array}
\right)
\label{ax:boundP-UA} \\
  \boxPMTL{[0,\BETA{u}/\delta]}{\neg\TAU{u} \wedge \bigwedge_{p \in \pre{u}} \MU{p}}
  \quad\Rightarrow\quad
  \left(
  \begin{array}{c}
    \bigvee_{p \in \pre{u}}\diamondMTL{=1}{\neg\MU{p}}
  \\ \vee \\
%
  \bigvee_{p \in \pre{u}}
  \left( \begin{array}{c}
      \boxPMTL{[0,\BETA{u}/\delta]}{\EPSILON{p}} \ \Rightarrow\ \diamondMTL{=1}{\neg\EPSILON{p}} 
      \\ \wedge \\
      \boxPMTL{[0,\BETA{u}/\delta]}{\neg\EPSILON{p}} \ \Rightarrow\ \diamondMTL{=1}{\EPSILON{p}} 
    \end{array}
  \right) 
   \\ \vee \\
   \diamondMTL{=1}{\TAU{u}} 
 \end{array}
\right)
\label{ax:boundN-UA} \\
 \text{Syntactically the same as in }\frf{ax:trans24UA}
 \label{ax:trans2-UA}
\end{gather}
\end{scriptsize}

The straightforward under-approximation of \frf{ax:boundP4UA} and \frf{ax:boundN4UA} yields formulas which have been re-arranged to eliminate redundant terms.
In fact, the time bound $(0, \BETA{u})$ in the antecedent becomes $[0, \BETA{u}/\delta]$ when under-approximated.
Hence, formulas such as $\boxPMTL{(0,\BETA{u})}{\gamma} \Rightarrow \neg\gamma \vee \nowonMTL{\neg\gamma}$ are under-approximated as $\boxPMTL{[0,\BETA{u}/\delta]}{\gamma} \Rightarrow \neg\gamma \vee \diamondMTL{[0,1]}{\neg\gamma}$.
However, $\neg\gamma$ never holds at the current instant because it would contradict the antecedent.
Correspondingly, such formulas can be simplified to $\boxPMTL{[0,\BETA{u}/\delta]}{\gamma} \Rightarrow \diamondMTL{=1}{\neg\gamma}$.

\subsubsection{Zero-time unmarking}
\begin{gather}
  \text{Syntactically the same as in \frf{ax:unm4UA}}
  \label{ax:unm-UA}
\end{gather}

\subsubsection{Initialization}
\begin{gather}
  \text{at $0$: } 
  \quad \bigwedge_{p \in P}\neg\MU{p}
  \wedge \diamondMTL{[1,2]}{\bigwedge_{p \in M_0} \MU{p}} 
  \ \wedge \ \bigwedge_{p \in P} \EPSILON{p} \wedge \bigwedge_{u \in T} \TAU{u}
\label{ax:init-UA}
\end{gather}

\subsection{Axiomatization for over-approximation} \label{sec:4overap}
Continuous-time operator $\becomesMTL{}$ becomes\footnote{After some semantic-preserving simplifications.} discrete-time operator $\becomesLMTL{}$ under over-ap\-prox\-i\-ma\-tion when it occurs to the left-hand side of implications, hence is suitable to describe antecedents of transitions that will be over-approximated.
However, the over-approximation of the same operator takes a different form in the right-hand side of implications.
In such cases, the over-approximation of formulas such as $\becomesMTL{x}$ is $\boxPMTL{[0,1]}{\neg x} \wedge \boxMTL{[0,1]}{x}$ which is clearly unsatisfiable.
Correspondingly, the whole over-approximation formulas would be unsatisfiable only for false antecedents, i.e., when no transition ever occurs.

After careful experimentation, we found that a workaround to this problem should exploit a weakening of the $\becomesMTL{}$ operators that occur in consequent formulas.
Let us illustrate the idea as simply as possible for two propositions $x,y$ and the formula $\becomesMTL{x} \Rightarrow \becomesMTL{y}$: every transition of $x$ occurs concurrently with a transition of $y$.
The formula is relaxed into the weaker $\becomesMTL{x} \Rightarrow \uptonowstrMTL{\neg y} \wedge \boxMTL{=\delta}{x \Rightarrow y}$: every transition of $x$ also triggers a transition of $y$ sometime in the future, as long as $x$ still holds $\delta$ time units in the future.
The new formula is essentially equivalent to the original one for non-Berkeley behaviors for the following reasons.
First, $x$ must still hold $\delta$ time units in the future, because its behavior is non-Berkeley for $\delta$; hence $y$ holds as well there and must transition somewhere over the interval $(0,\delta)$ from the current instant.
In addition, the transition of $y$ cannot occur asynchronously to the transition of $x$; otherwise two distinct transitions would occur within $\delta$ time units, against the non-Berkeleyness assumption.
In all, the two formulations are equivalent over non-Berkeley continuous time.
Correspondingly, the $\triggerOMTL{}$ operator is introduced and used in the right-hand side of implications in the following continuous-time formulas \fsrf{ax:u2m4OA_i}{ax:init4OA}.

\subsubsection{Places}

\begin{scriptsize}
\begin{gather}
  p \in M_0:
  \becomesMTL{\MU{p}} \ \Rightarrow\
  \left(
  \begin{array}{c}
  \bigvee_{u \in \pre{p}} 
  \left(
    \triggerOMTL{\MU{p} \leadsto \TAU{u}}
    \;\wedge\;
    \bigwedge_{u'\neq u \in \pre{p}} \triggerNOMTL{\MU{p} \leadsto \TAU{u'}}
  \right) 
  \\ \wedge \\
  \bigwedge_{u \in \post{p}} \triggerNOMTL{\MU{p} \leadsto \TAU{u}}
  \\ \vee \\
  \boxPMTL{[\delta, \infty)}{\neg\MU{p}}
  \end{array} \right)
 \label{ax:u2m4OA_i} \\
  p \notin M_0:
  \becomesMTL{\MU{p}} \ \Rightarrow\ 
  \left( \begin{array}{c}
  \bigvee_{u \in \pre{p}} 
  \left(
    \triggerOMTL{\MU{p} \leadsto \TAU{u}}
    \;\wedge\;
    \bigwedge_{u'\neq u \in \pre{p}} \triggerNOMTL{\MU{p} \leadsto \TAU{u'}}
  \right) 
  \\ \wedge \\
  \bigwedge_{u \in \post{p}} \triggerNOMTL{\MU{p} \leadsto \TAU{u}}
  \end{array} \right)
 \label{ax:u2m4OA} \\
 \becomesMTL{\neg\MU{p}} \quad\Rightarrow\quad
  \bigvee_{u \in \post{p}} 
  \left(
    \triggerOMTL{\neg\MU{p} \leadsto \TAU{u}}
    \;\wedge\;
    \bigwedge_{u'\neq u \in \post{p}} \triggerNOMTL{\neg\MU{p} \leadsto \TAU{u'}}
  \right) \wedge
  \bigwedge_{u \in \pre{p}} \triggerNOMTL{\neg\MU{p} \leadsto \TAU{u}}
  \label{ax:m2u4OA}
\end{gather}
\end{scriptsize}

\subsubsection{Transitions}

\begin{scriptsize}
\begin{gather}
  \triggerMTL{\TAU{u}} \quad\Rightarrow\quad
  \bigwedge_{p \in \pre{u}}
  \left( 
    \begin{array}{c}
      \uptonowstrMTL{\MU{p} \wedge \EPSILON{p}} \wedge \boxPMTL{[\delta,\ALPHA{u})}{ \MU{p} \wedge \EPSILON{p}}
      \\ \vee \\
      \uptonowstrMTL{\MU{p} \wedge \neg \EPSILON{p}} \wedge \boxPMTL{[\delta,\ALPHA{u})}{ \MU{p} \wedge \neg\EPSILON{p}}
    \end{array} 
  \right)
   \label{ax:enabling4OA} \\
\text{Same as }\frf{ax:boundP} \label{ax:boundP4OA} \\
\text{Same as }\frf{ax:boundN} \label{ax:boundN4OA} \\
  \becomesMTL{\TAU{u}} \quad\Rightarrow\quad
  \bigwedge_{p \in \pre{u}}
  \left( \begin{array}{c}
      \left( \begin{array}{c}
          \uptonowstrMTL{\MU{p}} \\\wedge\\ \boxMTL{=\delta}{\TAU{u} \Rightarrow \neg \MU{p}}
        \end{array}\right)
      \\ \vee \\
      \triggerOMTL{\TAU{u} \leadsto \EPSILON{p}} 
    \end{array} \right)
  \;\wedge\;
  \bigwedge_{p \in \post{u}}
  \left( \begin{array}{c}
      \left( \begin{array}{c}
          \uptonowstrMTL{\neg \MU{p}} \\\wedge\\ \boxMTL{=\delta}{\TAU{u} \Rightarrow \MU{p}}
        \end{array}\right)
      \\ \vee \\
      \triggerOMTL{\TAU{u} \leadsto \EPSILON{p}} 
    \end{array} \right)
  \nonumber \\
  \becomesMTL{\neg\TAU{u}} \quad\Rightarrow\quad
  \bigwedge_{p \in \pre{u}}
  \left( \begin{array}{c}
      \left( \begin{array}{c}
         \uptonowstrMTL{\MU{p}} \\\wedge\\ \boxMTL{=\delta}{\neg\TAU{u} \Rightarrow \neg \MU{p}}
        \end{array}\right)
      \\ \vee \\
      \triggerOMTL{\neg\TAU{u} \leadsto \EPSILON{p}} 
    \end{array} \right)
  \;\wedge\;
  \bigwedge_{p \in \post{u}}
  \left( \begin{array}{c}
      \left( \begin{array}{c}
          \uptonowstrMTL{\neg \MU{p}} \\\wedge\\ \boxMTL{=\delta}{\neg\TAU{u} \Rightarrow \MU{p}}
        \end{array}\right)
      \\ \vee \\
      \triggerOMTL{\neg\TAU{u} \leadsto \EPSILON{p}} 
    \end{array} \right) 
\label{ax:trans24OA}
\end{gather}
\end{scriptsize}

\subsubsection{Zero-time unmarking}

\begin{scriptsize}
\begin{gather}
  \becomesMTL{\EPSILON{p}} \quad\Rightarrow\quad
  \bigvee_{\substack{u_a \in \pre{p} \\ u_b \in \post{p}}}
  \left( \begin{array}{c}
      \triggerOMTL{\EPSILON{p} \leadsto \TAU{u_a}}
      \;\wedge\;
      \bigwedge_{u'\neq u_a \in \pre{p}} \triggerNOMTL{\EPSILON{p} \leadsto \TAU{u'}}   
      \\ \wedge \\
      \triggerOMTL{\EPSILON{p} \leadsto \TAU{u_b}}
      \;\wedge\;
      \bigwedge_{u'\neq u_b \in \post{p}} \triggerNOMTL{\EPSILON{p} \leadsto \TAU{u'}}
    \end{array} \right) 
\nonumber \\
  \becomesMTL{\neg \EPSILON{p}} \quad\Rightarrow\quad
  \bigvee_{\substack{u_a \in \pre{p} \\ u_b \in \post{p}}}
  \left( \begin{array}{c}
      \triggerOMTL{\neg \EPSILON{p} \leadsto \TAU{u_a}}
      \;\wedge\;
      \bigwedge_{u'\neq u_a \in \pre{p}} \triggerNOMTL{\neg \EPSILON{p} \leadsto \TAU{u'}}   
      \\ \wedge \\
      \triggerOMTL{\neg \EPSILON{p} \leadsto \TAU{u_b}}
      \;\wedge\;
      \bigwedge_{u'\neq u_b \in \post{p}} \triggerNOMTL{\neg \EPSILON{p} \leadsto \TAU{u'}}
    \end{array} \right) 
\label{ax:unm4OA}
\end{gather}
\end{scriptsize}

\subsubsection{Initialization}
\begin{gather}
  \boxPMTL{(0, \infty)}{\perp} 
  \quad\Rightarrow\quad 
  \bigwedge_{p \in P}\neg\MU{p}
  \wedge \diamondMTL{[0,2\delta]}{\bigwedge_{p \in M_0} \MU{p}} 
  \ \wedge \ \nowonMTL{\bigwedge_{p \in P} \EPSILON{p} \wedge \bigwedge_{u \in T} \TAU{u}}
\label{ax:init4OA}
\end{gather}

The observations that have been introduced at the beginning of this section can be leveraged to provide a rigorous proof that \fsrf{ax:u2m4OA_i}{ax:init4OA} are equivalent to the original \fsrf{ax:u2m_i}{ax:init} over non-Berkeley continuous time.
We omit the details for brevity.

\subsection{Over-approximation} \label{sec:OA}
The over-approximations of \fsrf{ax:u2m4OA_i}{ax:init4OA} are reported as formulas \fsrf{ax:u2m-OA_i}{ax:init-OA}.
Notice the lower- and upper-bound relaxations in \fsrf{ax:enabling-OA}{ax:boundN-OA}, in accordance with the notion of over-ap\-prox\-i\-ma\-tion.

\subsubsection{Places}

\begin{scriptsize}
\begin{gather}
  p \in M_0:
  \becomesLMTL{\MU{p}} \ \Rightarrow\ 
  \left( \begin{array}{c}
  \bigvee_{u \in \pre{p}} 
  \left(
    \triggerOMTL{\MU{p} \leadsto \TAU{u}}
    \;\wedge\;
    \bigwedge_{u'\neq u \in \pre{p}} \triggerNOMTL{\MU{p} \leadsto \TAU{u'}}
  \right) 
  \\ \wedge \\
  \bigwedge_{u \in \post{p}} \triggerNOMTL{\MU{p} \leadsto \TAU{u}}
  \\ \vee \\
  \boxPMTL{[\delta, \infty)}{\neg\MU{p}}
  \end{array} \right)
 \label{ax:u2m-OA_i} \\
  p \notin M_0:
  \becomesLMTL{\MU{p}} \ \Rightarrow\ 
  \left( \begin{array}{c}
  \bigvee_{u \in \pre{p}} 
  \left(
    \triggerOMTL{\MU{p} \leadsto \TAU{u}}
    \;\wedge\;
    \bigwedge_{u'\neq u \in \pre{p}} \triggerNOMTL{\MU{p} \leadsto \TAU{u'}}
  \right) 
  \\ \wedge \\
  \bigwedge_{u \in \post{p}} \triggerNOMTL{\MU{p} \leadsto \TAU{u}}
  \end{array} \right)
 \label{ax:u2m-OA} \\
 \becomesLMTL{\neg\MU{p}} \quad\Rightarrow\quad
  \bigvee_{u \in \post{p}} 
  \left(
    \triggerOMTL{\neg\MU{p} \leadsto \TAU{u}}
    \;\wedge\;
    \bigwedge_{u'\neq u \in \post{p}} \triggerNOMTL{\neg\MU{p} \leadsto \TAU{u'}}
  \right) \wedge
  \bigwedge_{u \in \pre{p}} \triggerNOMTL{\neg\MU{p} \leadsto \TAU{u}}
  \label{ax:m2u-OA}
\end{gather}
\end{scriptsize}

\subsubsection{Transitions}

\begin{scriptsize}
\begin{gather}
  \triggerMTL{\TAU{u}} \quad\Rightarrow\quad
  \bigwedge_{p \in \pre{u}}
  \left( 
    \begin{array}{c}
      \boxPMTL{[0,\ALPHA{u}/\delta+1]}{\MU{p} \wedge \EPSILON{p}}
      \\ \vee \\
      \boxPMTL{[0,\ALPHA{u}/\delta+1]}{\MU{p} \wedge \neg\EPSILON{p}}
    \end{array} 
  \right)
   \label{ax:enabling-OA} \\
  \boxPMTL{[1,\BETA{u}/\delta-1]}{\TAU{u} \wedge \bigwedge_{p \in \pre{u}} \MU{p}}
  \Rightarrow
  \left(
  \begin{array}{c}
    \bigvee_{p \in \pre{u}}(\neg\MU{p} \vee \boxMTL{[0,1]}{\neg\MU{p}})
  \\ \vee \\
%
  \bigvee_{p \in \pre{u}}
  \left( \begin{array}{c}
      \boxPMTL{[1,\BETA{u}/\delta-1]}{\EPSILON{p}} \ \Rightarrow\ \neg\EPSILON{p} \vee \boxMTL{[0,1]}{\neg\EPSILON{p}} 
      \\ \wedge \\
      \boxPMTL{[0,\BETA{u}/\delta-1]}{\neg\EPSILON{p}} \ \Rightarrow\ \EPSILON{p} \vee \diamondMTL{[0,1]}{\EPSILON{p}} 
    \end{array}
  \right) 
   \\ \vee \\
%
       \neg \TAU{u} 
 \end{array}
\right)
\label{ax:boundP-OA} \\
  \boxPMTL{[1,\BETA{u}/\delta-1]}{\neg\TAU{u} \wedge \bigwedge_{p \in \pre{u}} \MU{p}}
  \Rightarrow
  \left(
  \begin{array}{c}
    \bigvee_{p \in \pre{u}}\boxMTL{[0,1]}{\neg\MU{p}}
  \\ \vee \\
%
  \bigvee_{p \in \pre{u}}
  \left( \begin{array}{c}
      \boxPMTL{[1,\BETA{u}/\delta-1]}{\EPSILON{p}} \ \Rightarrow\ \neg\EPSILON{p} \vee \boxMTL{[0,1]}{\neg\EPSILON{p}} 
      \\ \wedge \\
      \boxPMTL{[0,\BETA{u}/\delta-1]}{\neg\EPSILON{p}} \ \Rightarrow\ \EPSILON{p} \vee \diamondMTL{[0,1]}{\EPSILON{p}} 
    \end{array}
  \right) 
   \\ \vee \\
%
       \TAU{u} 
 \end{array}
\right)
\label{ax:boundN-OA} \\
  \becomesLMTL{\TAU{u}} \quad\Rightarrow\quad
  \bigwedge_{p \in \pre{u}}
  \left( \begin{array}{c}
      \left( \begin{array}{c}
          \boxPMTL{[0,1]}{\MU{p}} \\\wedge\\ \boxMTL{[0,2]}{\TAU{u} \Rightarrow \neg \MU{p}}
        \end{array}\right)
      \\ \vee \\
      \triggerOMTL{\TAU{u} \leadsto \EPSILON{p}} 
    \end{array} \right)
  \;\wedge\;
  \bigwedge_{p \in \post{u}}
  \left( \begin{array}{c}
      \left( \begin{array}{c}
          \boxPMTL{[0,1]}{\neg \MU{p}} \\\wedge\\ \boxMTL{[0,2]}{\TAU{u} \Rightarrow \MU{p}}
        \end{array}\right)
      \\ \vee \\
      \triggerOMTL{\TAU{u} \leadsto \EPSILON{p}} 
    \end{array} \right)
  \nonumber \\
  \becomesLMTL{\neg\TAU{u}} \quad\Rightarrow\quad
  \bigwedge_{p \in \pre{u}}
  \left( \begin{array}{c}
      \left( \begin{array}{c}
         \boxPMTL{[0,1]}{\MU{p}} \\\wedge\\ \boxMTL{[0,2]}{\neg\TAU{u} \Rightarrow \neg \MU{p}}
        \end{array}\right)
      \\ \vee \\
      \triggerOMTL{\neg\TAU{u} \leadsto \EPSILON{p}} 
    \end{array} \right)
  \;\wedge\;
  \bigwedge_{p \in \post{u}}
  \left( \begin{array}{c}
      \left( \begin{array}{c}
          \boxPMTL{[0,1]}{\neg \MU{p}} \\\wedge\\ \boxMTL{[0,2]}{\neg\TAU{u} \Rightarrow \MU{p}}
        \end{array}\right)
      \\ \vee \\
      \triggerOMTL{\neg\TAU{u} \leadsto \EPSILON{p}} 
    \end{array} \right) 
\label{ax:trans2-OA}
\end{gather}
\end{scriptsize}

Similarly as with under-approximation, formulas have been conveniently simplified: the term $\uptonowstrMTL{\MU{p} \wedge \EPSILON{p}}$ in the consequent of \frf{ax:enabling4OA} is over-approximated to \linebreak $\boxPMTL{[0,1]}{\MU{p} \wedge \EPSILON{p}}$, which is subsumed by the other term $\boxPMTL{[0, \ALPHA{u}/ \delta +1 )}{\MU{p} \wedge \EPSILON{p}}$ in the over-approximation. (In fact, $\ALPHA{u}/\delta+1 \geq 2$ is the case).
Subformulas $\neg \TAU{u} \vee \boxMTL{[0,1]}{\neg\TAU{u}}$ and $\TAU{u} \vee \boxMTL{[0,1]}{\TAU{u}}$ in the over-approximations \frf{ax:boundN-OA} and \frf{ax:boundN-OA}, respectively, can also be simplified.
In fact, \frf{ax:enabling-OA} enforces marking and no zero-time unmarking for at least 3 time units whenever $\tau_u$ is triggered; hence $\MU{p}$ cannot be triggered over $[0,1]$ so that the terms $\boxMTL{[0,1]}{\neg\TAU{u}}$ and $\boxMTL{[0,1]}{\TAU{u}}$ are redundant.

\subsubsection{Zero-time unmarking}

\begin{scriptsize}
\begin{gather}
  \becomesLMTL{\EPSILON{p}} \quad\Rightarrow\quad
  \bigvee_{\substack{u_a \in \pre{p} \\ u_b \in \post{p}}}
  \left( \begin{array}{c}
      \triggerOMTL{\EPSILON{p} \leadsto \TAU{u_a}}
      \;\wedge\;
      \bigwedge_{u'\neq u_a \in \pre{p}} \triggerNOMTL{\EPSILON{p} \leadsto \TAU{u'}}   
      \\ \wedge \\
      \triggerOMTL{\EPSILON{p} \leadsto \TAU{u_b}}
      \;\wedge\;
      \bigwedge_{u'\neq u_b \in \post{p}} \triggerNOMTL{\EPSILON{p} \leadsto \TAU{u'}}
    \end{array} \right) 
\nonumber \\
  \becomesLMTL{\neg \EPSILON{p}} \quad\Rightarrow\quad
  \bigvee_{\substack{u_a \in \pre{p} \\ u_b \in \post{p}}}
  \left( \begin{array}{c}
      \triggerOMTL{\neg \EPSILON{p} \leadsto \TAU{u_a}}
      \;\wedge\;
      \bigwedge_{u'\neq u_a \in \pre{p}} \triggerNOMTL{\neg \EPSILON{p} \leadsto \TAU{u'}}   
      \\ \wedge \\
      \triggerOMTL{\neg \EPSILON{p} \leadsto \TAU{u_b}}
      \;\wedge\;
      \bigwedge_{u'\neq u_b \in \post{p}} \triggerNOMTL{\neg \EPSILON{p} \leadsto \TAU{u'}}
    \end{array} \right) 
\label{ax:unm-OA}
\end{gather}
\end{scriptsize}

\subsubsection{Initialization}
\begin{gather}
 \text{at $0$: } 
  \quad \bigwedge_{p \in P}\neg\MU{p}
  \wedge \diamondMTL{=1}{\bigwedge_{p \in M_0} \MU{p}} 
  \ \wedge \ \boxMTL{[0,1]}{\bigwedge_{p \in P} \EPSILON{p} \wedge \bigwedge_{u \in T} \TAU{u}}
\label{ax:init-OA}
\end{gather}

\subsection{Quality of discrete-time approximations}
Proposition \ref{prop:approximations} guarantees that under-ap\-prox\-i\-ma\-tions preserve validity and over-ap\-prox\-i\-ma\-tions preserve counterexamples.
It does not say anything about the \emph{quality} (or completeness) of such approximations; in particular an under-approximation can preserve validity trivially by being contradictory (i.e., inconsistent), and an over-approximation can preserve counterexamples trivially by being identically valid.

In order to make sure this is not the case, let us introduce a set of constraints that guarantees no degenerate behaviors are modeled in the approximations.
Consider formulas involving metric intervals, namely \fsrf{ax:enabling-UA}{ax:boundN-UA} for the under-approximations and \fsrf{ax:enabling-OA}{ax:boundN-OA} for the over-approximation.
We should check that, for every transition $u$ with dense-time firing interval $[\ALPHA{u}, \BETA{u}]$:
\begin{itemize}
  \item \emph {non-emptiness.} Metric intervals are non-empty; that is $\ALPHA{u} \geq 3\delta$ from the under-approximation and $\ALPHA{u} \geq -\delta$, $\BETA{u} \geq 2\delta$ from the over-approximation.

  \item \emph{consistency.} The the minimum enabling interval (defined in \frf{ax:enabling-UA} and \frf{ax:enabling-OA} for under- and over-approximation respectively) is smaller than the maximum enabling interval (defined in \fsrf{ax:boundP-UA}{ax:boundN-UA} and \fsrf{ax:boundP-OA}{ax:boundN-OA} for under- and over-approximation respectively).
    Correspondingly, we have the constraints $\BETA{u} \geq \ALPHA{u} - 2\delta$ from the under-approximation and $\BETA{u} \geq \ALPHA{u} + 2\delta$ from the over-approximation.
\end{itemize}

The constraints can be summarized as $\ALPHA{u} \geq 3\delta$ and  $\BETA{u} \geq \ALPHA{u} + 2\delta$.
In our examples, we will consider only non-degenerate TPN satisfying these constraints.


%% file: CaseStudy.tex
\section{Multi-Paradigm Modeling and Verification at Work}\label{sec:casestudy}

The multi-paradigm modeling technique presented in this paper is supported by the \zot{} bounded satisfiability checker \cite{zot,PMS07}.
More precisely, we exploited the flexibility provided by the SAT-based approach pursued by \zot{}, and implemented several separate plugins to deal with the various allowed formalisms.
In particular, the tool now includes plugins capable of dealing with dense-time MTL formulas \cite{FPR08-FM08}, with timed automata \cite{FPR08-ICFEM08}, and with timed Petri nets (using the formalization presented in Section \ref{sec:TPN}).
In addition, \zot{} is natively capable of accepting discrete-time MTL formulas as input language.
The plugins provide primitives through which the user can define the system to be analyzed as a mixture of timed automata, dense- and discrete-time MTL formulas, and timed Petri nets.
The properties to be verified for the system can also be described as a combination of fragments written using the aforementioned formal languages, though they are usually formalized through MTL formulas (either using dense or discrete time).

The tool then automatically builds, for the dense-time fragments of the system and of the property to be analyzed, the two discrete-time approximation formulas of Proposition \ref{prop:approximations}.
These formulas, in possibly conjunction with MTL formulas natively written using a discrete notion of time, are checked for validity over time $\naturals$; the results of the validity check allows one to infer the validity of the integrated model, according to Proposition \ref{prop:approximations}.

The multi-paradigm verification process in \zot{} consists of three sequential phases.
First, the discrete-time MTL formulas of Proposition \ref{prop:approximations} are built and are translated into a propositional satisfiability (SAT) problem.
Second, the SAT instance (possibly including MTL formulas directly written using a discrete notion of time) is put into conjunctive normal form (CNF), a standard input format for SAT solvers.
Third, the CNF formula is fed to a SAT solving engine (such as MiniSat, zChaff, or MiraXT).

\subsection{An Example of Multi-paradigm Modeling and Verification}
We demonstrate how the modeling and verification technique presented in this paper works in practice through an example consisting of a fragment of a realistic monitoring system, which could be part of a larger supervision and control system.

The monitoring subsystem is composed of three identical sensors, a middle component that is in charge of acquiring and pre-processing the data from the sensors, and a data management component that further elaborates the data (e.g., to select appropriate control actions).
%
%
%
For reasons of dependability (by redundancy), the three sensors measure the same quantity (whose nature is of no relevance in this example).
Each one of them  senses independently the measured quantity at a certain rate which is in general aperiodic; however, while the acquisition rate can vary, the distance between consecutive acquisitions must always be no less than $T/2$ and no more than $T$ time units.
Each sensor keeps track of only the last measurement, hence every new sensed value replaces the one stored by the sensor.

The data acquisition component retrieves data from the three sensors in a ``pull'' fashion.
More precisely, when all three sensors have a fresh measurement available, with a delay of at least $T/10$ units, but of no more than $T/5$ time units, the data acquisition component collects the three values from the sensors (which then become stale, as they have been acquired).
After having retrieved the three measurements, the component processes them (e.g., it computes a derived measurement as the average of the sensed values); the process takes between $T/5$ and $T/2$ time units.

After having computed the derived measurement, the data acquisition component sends it to the data manager, this time using a ``push'' policy which requires an acknowledgement of the data reception by the latter.
The data acquisition component tries to send data to the data manager at most twice.
If both attempts at data transmission fail (for example because a timeout for the reception acknowledgement by the data manager expires, or because the latter signals a reception error), the data transmission terminates with an error.

\begin{figure}[!htbp]
  \centering
  \includegraphics[width=12cm]{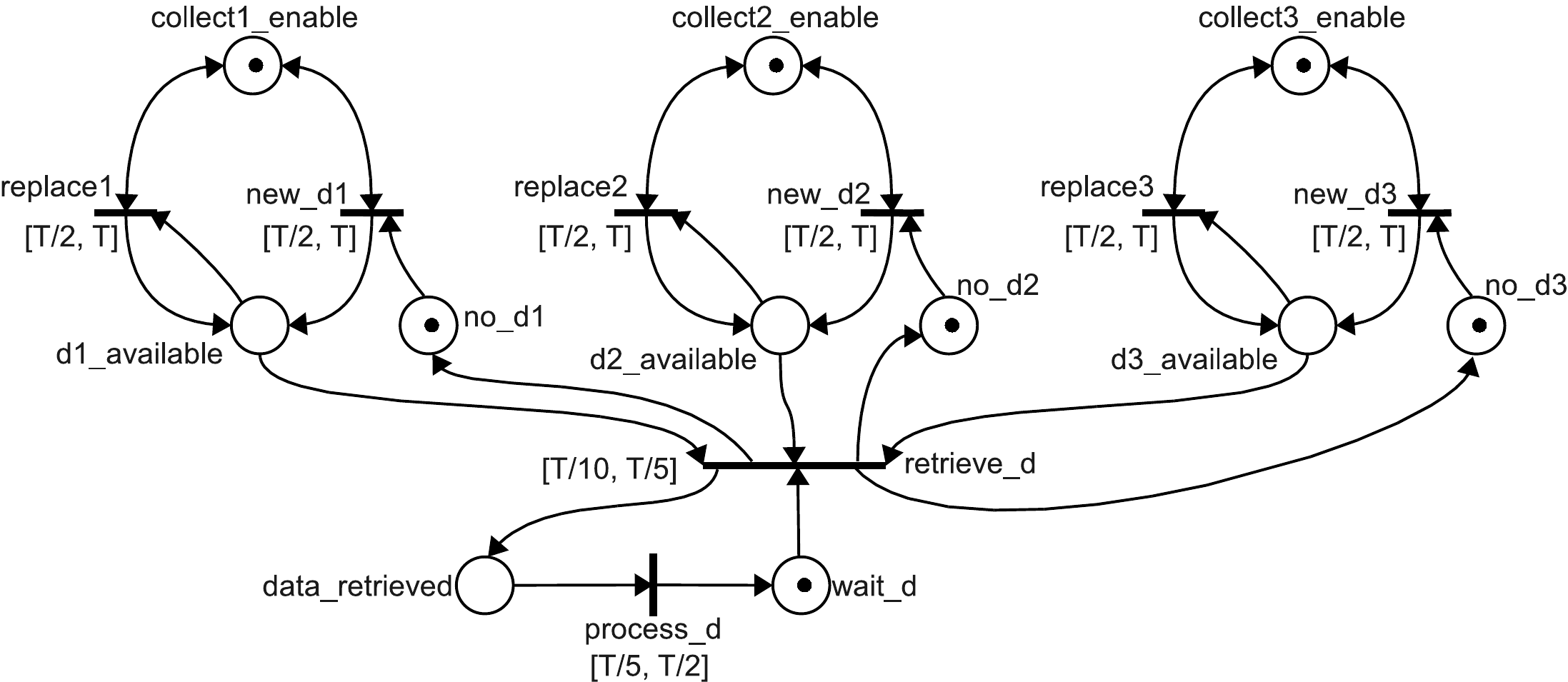}
  \caption{Fragment of monitoring system modeled through a timed Petri net.}
  \label{fig:TPN}
\end{figure}

First, we model the mechanism through which the three sensors collect data from the field and the data acquisition component retrieves them for the pre-processing phase.
This fragment of the model is described through a timed Petri net, and is depicted in Figure \ref{fig:TPN}.

%

In a multiple-paradigm framework, the reasons that lead to the choice of a notation instead of another often include a certain degree of arbitrariness.
In this case, however, we chose to model the data acquisition part of the system through a TPN since we felt that the inherent asynchrony with which the three sensors collect data from the field was naturally matched by the asynchronous nature of a TPN and its tokens \cite{FMMR09}.
While it is undeniable that different modelers might have made different choices, we maintain that TPN are well-suited (although not necessarily indispensable) in this case.

\begin{figure}[!htbp]
  \centering
  \includegraphics[width=12cm]{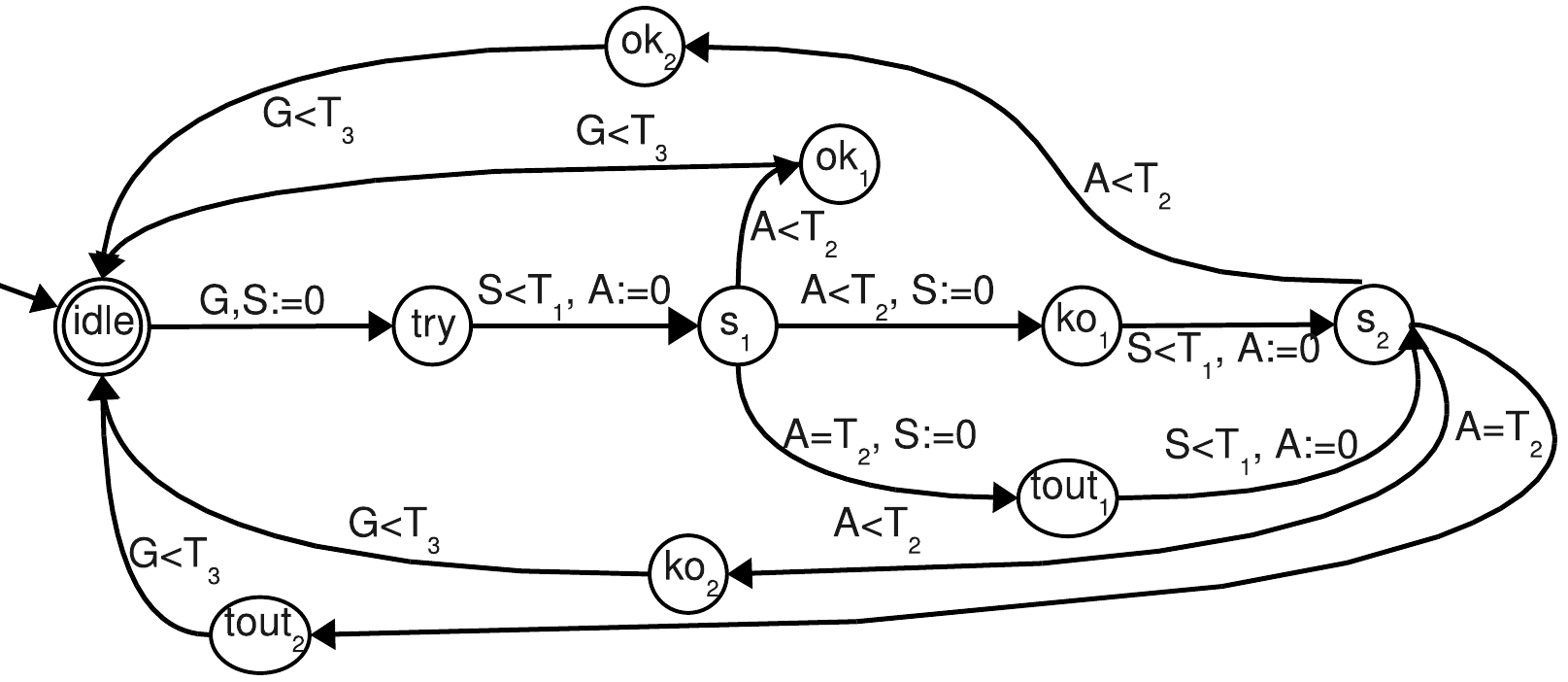}
  \caption{Fragment of data acquisition system modeled through a timed automaton.}
  \label{fig:TA}
\end{figure}

A further fragment of the formal model of the monitoring system is shown in Figure \ref{fig:TA}.
It represents, through the formalism of timed automata presented in \cite{FPR08-ICFEM08}, the transmission protocol that the data acquisition component uses to send refined values to the data manager.\footnote{As remarked in \cite{FPR08-ICFEM08}, since, in our formalization, the definition of clock constraints forbids the introduction of exact constraints such as $A = T_2$, such constraints represent a shorthand for the valid clock constraint $T_2 \leq A < T + \delta$.}

For this second fragment of the system, the formalism of timed automata was chosen, with a certain degree of arbitrariness, because it was deemed capable of representing the timing constraints on the protocol in a more natural way, especially for what concerns the constraint on the overall duration of the process.


Finally, MTL formulas are added to ``bridge the gap'' between the fragments shown in Figures \ref{fig:TPN} and \ref{fig:TA}.
This is achieved by the two following formulas, which define, respectively, that the transmission procedure can begin only if a pre-processed measurement value has been produced by the data acquisition component in the last $T$ time units (\ref{eq:CNtrans}) and if the system is not in the middle of a data transmission (i.e., it is $\idle$), and a new datum is being processed, a transmission will start within $T/2$ time units, due to the upper bound of \textit{process\_d} transition (\ref{eq:CStrans}).

\begin{eqnarray}
     \try \quad\Rightarrow\quad  \diamondPMTL{(0,T/2]}{\dataret} \label{eq:CNtrans}\\
     \dataret \land \idle \quad\Rightarrow\quad  \diamondMTL{(0,T/2]}{\try} \label{eq:CStrans}
\end{eqnarray}

Notice that the automata of Figures \ref{fig:TPN} and \ref{fig:TA} are defined, as per the formalizations of \cite{FPR08-ICFEM08} and of Section \ref{sec:TPN}, over a continuous notion of time.
This choice for the time domain of these two system fragments is justified by the fact that they deal with parts of the system interacting with physical elements (measured quantities, transmission channel), for which a continuous time seems better suited.

Formulas (\ref{eq:CNtrans}) and (\ref{eq:CStrans}), instead, describe a software synchronization mechanism within the application.
As a consequence, discrete time is more suitable to describe this part of the system, hence formulas (\ref{eq:CNtrans}) and (\ref{eq:CStrans}) are to be interpreted accordingly.

Finally, the model of the system to be verified is built by conjoining the discrete-time approximations for the fragments of Figures \ref{fig:TPN}-\ref{fig:TA} and the discrete-time MTL formulas (\ref{eq:CNtrans})-(\ref{eq:CStrans}).
More precisely, if $\psi^{\underapempty}_{N}$ and $\psi^{\overapempty}_{N}$ are the continuous-time MTL formulas capturing the semantics of the net of Figure \ref{fig:TPN} (see Section \ref{sec:TPN}), $\psi^{\underapempty}_{A}$, $\psi^{\overapempty}_{A}$ are the continuous-time MTL formulas for the automaton of Figure \ref{fig:TA}, $\psi_{L}$ is the discrete-time formula $\psi_{L} = (\ref{eq:CNtrans}) \land (\ref{eq:CStrans})$, and $\prop$ is the continuous-time property to be checked for the system, then we have:

\begin{footnotesize}
$$
\begin{array}{l}
 \phi^+ = \Alw{\underap{\psi^{\underapempty}_{N}} \land \underap{\psi^{\underapempty}_{A}} \land \psi_{L}} \Rightarrow \Alw{\overap{\prop}} \\
 \phi^- = \Alw{\overap{\psi^{\overapempty}_{N}} \land \overap{\psi^{\overapempty}_{A}} \land \psi_{L}} \Rightarrow \Alw{\underap{\prop}}
\end{array}
$$
\end{footnotesize}

Note that formula $\psi_L$, which is to be interpreted over discrete time, must not be approximated.
Then, if $\phi^+$ is $\naturals$-valid, we can draw some interesting conclusions.

First, if one implements a continuous-time system that does not vary faster than the sampling time $\delta$ (i.e., whose behaviors are in $\Bchi^\delta$), which satisfies $\psi_N$, $\psi_A$, and a continuous-time MTL formula $\psi'$ such that $\underap{\psi'_L} = \psi_L$, then property $\prop$ holds for this system.

It can be shown that, for any continuous-time MTL formula $\phi$, the set of behaviors satisfying $\overap{\phi}$ is a subset of those satisfying $\underap{\phi}$ (i.e., $\{b \mid b \modelstime{\naturals} \overap{\phi}\} \subseteq \{b \mid b \modelstime{\naturals} \underap{\phi}\}$).
In addition, given a discrete-time behavior $b$ that satisfies $\overap{\phi}$, from \cite[Lemma 3]{FPR08-FM08} we have that any continuous-time non-Berkeley behavior $b'$ for which $b$ is a sampling satisfies $\phi$.
Then, any way one reconstructs a continuous-time non-Berkeley behavior $b'$ from a discrete-time one that satisfies $\overap{\phi}$, $b'$ satisfies $\phi$. This leads us to conclude that, if one builds a discrete-time system (e.g., a piece of software) which implements --- that is, satisfies --- $\overap{\psi^{\overapempty}_{N}}$, $\overap{\psi^{\overapempty}_{A}}$, $\psi_L$, this satisfies discrete-time property $\overap{\prop}$; in addition, any way one uses a discrete-time behavior of this system to reconstruct a continuous-time, non-Berkeley behavior, the latter satisfies $\psi_N$, $\psi_A$, and $\prop$.

Finally, if $\phi^-$ is not $\naturals$-valid, a discrete-time system implementing $\overap{\psi^{\overapempty}_{N}}$, $\overap{\psi^{\overapempty}_{A}}$, $\psi_L$ violates property $\underap{\prop}$.

\paragraph{Verification.}
We used the system model presented above to check a number of properties to validate the effectiveness of our approach.
Table \ref{tab:exp_res} shows the results, and duration of the tests.
More precisely, for each test the table reports: the checked property; the values of the timing parameters in the model (i.e., $T_1, T_2, T_3$, $T$); the temporal bound $k$ of the time domain (as \zot{} is a bounded satisfiability checker, it considers all the behaviors with period $\le k$); the total amount of time to perform each phase of the verification, namely formula building (including transformation into conjunctive normal form), and propositional satisfiability checking; the results of the tests; the size (in millions of clauses) of the formula fed to the SAT-solver.\footnote{The verification tool and the complete model used for verification can be found at \texttt{http://home.dei.polimi.it/pradella}. Tests have been performed on a PC equipped with two Intel Xeon E5335 Quad-Core Processor 2GHz, 16 Gb of RAM, and  GNU/Linux (kernel 2.6.29), using a single core for each test. \zot{} used the SAT-solver MiniSat 2.}
Tests were performed instantiating the parameters with different values to get an idea of how the performance of the verification algorithm is affected, both in terms of time to complete the verification and of whether the verification attempt is conclusive.
In addition, the timed interaction between the data acquisition and monitoring subsystems is quite subtle and the properties under verification hold in every run of the system only for certain combinations of parameter values.
Automated verification allowed us to investigate this fact in some detail.

First, we checked some properties concerning the liveness of the data collection by a sensor $X$ (with $X \in \{1, 2, 3\}$).
More precisely, we analyzed whether property (\ref{p2a}) holds for the model.\footnote{Recall that all properties to be proved are implicitly closed with the $\Alw{}$ operator.}

\begin{equation} \label{p2a}
       \begin{array}{c}
       \tait{replaceX} \land \tait{new\_dX} \Rightarrow \\\diamondMTL{(0,T+\delta]}{\tait{replaceX} \land \neg\tait{new\_dX} \lor \neg\tait{replaceX} \land \tait{new\_dX}} \\ \land \\
       \tait{replaceX} \land \neg\tait{new\_dX} \Rightarrow \\\diamondMTL{(0,T+\delta]}{\tait{replaceX} \land \tait{new\_dX} \lor \neg\tait{replaceX} \land \neg\tait{new\_dX}}\\ \land \\
       \neg\tait{replaceX} \land \tait{new\_dX} \Rightarrow \\\diamondMTL{(0,T+\delta]}{\neg\tait{replaceX} \land \neg\tait{new\_dX} \lor \tait{replaceX} \land \tait{new\_dX}}\\ \land \\
       \neg\tait{replaceX} \land \neg\tait{new\_dX} \Rightarrow \\\diamondMTL{(0,T+\delta]}{\neg\tait{replaceX} \land \tait{new\_dX} \lor \tait{replaceX} \land \neg\tait{new\_dX}}
       \end{array}
\end{equation}

Formula (\ref{p2a}) states that triggering events of $\tait{replaceX}$ and $\tait{new\_dX}$ transitions must occur within $T+\delta$ (with $\delta$ the sampling period) time instants in the future, i.e., that either $\tait{replaceX}$ or $\tait{new\_dX}$ must change value within the next $T+\delta$ time instants.
The property does not hold in general, since a firing of transition $\tait{retrieve\_d}$ would reset the time counters for transitions $\tait{replaceX}$ and $\tait{new\_dX}$.
This fact can be pointed out by checking $\phi^-$, with $\prop = (\ref{p2a})$, which is unsatisfiable, as shown in Table \ref{tab:exp_res}.

If the additional hypothesis that transition $\tait{retrieve\_d}$ does not fire along $(0,T+\delta]$, \frf{p2a} can however be shown to hold.
More precisely, if (\ref{p2a}) is rewritten, as shown in formula (\ref{p2b}), by adding to the antecedents the condition that predicate $\tait{retrieve\_d}$ does not change in $(0,T+\delta]$ (i.e., transition $\tait{retrieve\_d}$ does not fire in that interval), then the new $\phi^+$ is $\naturals$-valid (as Table \ref{tab:exp_res} shows), hence (\ref{p2b}) holds for the system.
\begin{equation} \label{p2b}
       \begin{array}{c} 
         \boxMTL{(0,T+\delta]}{\tait{retrieve\_d}} \land \tait{replaceX} \land \tait{new\_dX} \Rightarrow \\ 
         \diamondMTL{(0,T+\delta]}{\tait{replaceX} \land \neg\tait{new\_dX} \lor \neg\tait{replaceX} \land \tait{new\_dX}} \\
         \wedge \dots \wedge \\ 
         \boxMTL{(0,T+\delta]}{\tait{retrieve\_d}} \land \neg\tait{replaceX} \land \neg\tait{new\_dX} \Rightarrow \\
         \diamondMTL{(0,T+\delta]}{\neg\tait{replaceX} \land \tait{new\_dX} \lor \neg\tait{replaceX} \land \tait{new\_dX}} \\
%
%
       \bigvee \\
%
         \boxMTL{(0,T+\delta]}{\neg \tait{retrieve\_d}} \land \tait{replaceX} \land \neg\tait{new\_dX} \Rightarrow \\
       \diamondMTL{(0,T+\delta]}{\tait{replaceX} \land \tait{new\_dX} \lor \neg\tait{replaceX} \land \neg\tait{new\_dX}} \\
       \wedge \dots \wedge \\
%
         \boxMTL{(0,T+\delta]}{\neg\tait{retrieve\_d}} \land \neg\tait{replaceX} \land \neg\tait{new\_dX} \Rightarrow \\
        \diamondMTL{(0,T+\delta]}{\neg\tait{replaceX} \land \tait{new\_dX} \lor \neg\tait{replaceX} \land \tait{new\_dX}}
       \end{array}
\end{equation}

Another liveness property is formalized by formula (\ref{p1a}), which states that a datum is retrieved (i.e., place $\dataret$ is marked) at least every $\frac{3T}{2}$ time units.

\begin{equation}\label{p1a}
     \diamondMTL{(0,\frac{3T}{2}]}{\dataret}
\end{equation}

Property (\ref{p1a}) cannot be established with our verification technique as it falls in the incompleteness region (i.e., $\phi^+$ is not valid and $\phi^-$ is valid, as Table \ref{tab:exp_res} shows); from the automated check we cannot draw a definitive conclusion on the validity of the property for the system.
If, however, the temporal bound of formula (\ref{p1a}) is slightly relaxed as in formula (\ref{p1c}), not only the verification is conclusive, but it shows that the property in fact holds for the system.


\begin{equation}\label{p1c}
     \diamondMTL{(0,2T]}{\dataret}
\end{equation}

Verification also shows that the original formula ($\ref{p1a}$) holds if the bound on transitions $\tait{replaceX}$ of the TPN is changed to $[\frac{4T}{5}, T]$ (property (\ref{p1a}') in Table \ref{tab:exp_res}).

Formula (\ref{p3}) expresses the maximum delay between sensor collect and data send.
More precisely, if each sensor has provided a measurement and transition $\tait{retrieve\_d}$ fires, then the timed automaton will enter state $\try$ within $T$ instants.
The validity of this formula would allow us to check that the two parts of the system modeled by the TPN and by the TA are correctly ``bridged'' by axioms (\ref{eq:CNtrans}) and (\ref{eq:CStrans}).
As Table \ref{tab:exp_res} shows, property (\ref{p3}) does not hold; this occurs because, when place $\tait{data\_retrieved}$ is marked, the TA might not be in state $\idle$.

\begin{equation}\label{p3}
  \tait{data\_retrieved} \Rightarrow \diamondMTL{(0, T]}{\tait{try}}
\end{equation}

Axiom (\ref{eq:CStrans}) states that a $\try$ state is entered within $T/2$ if $\tait{data\_retrieved}$ holds when $\idle$ holds.
Then, a deeper analysis on the timing constraints suggests that this condition depends on the maximum transmission time $T_3$ of the TA, which defines the maximum delay between two consecutive occurrences of $\idle$.
If the system is in $\tait{data\_retrieved}$ and not in $\idle$, then the next $\idle$ state will be within $T_3$ instants in the future; moreover, $\tait{data\_retrieved}$ will be unmarked within $T/2$.
This suggests that the following property (\ref{p3a}) is valid:

\begin{equation}\label{p3a}
  \boxMTL{(0,T_3]}{\tait{data\_retrieved}} \Rightarrow \diamondMTL{(0, T]}{\tait{try}}
\end{equation}

This property also falls in the incompleteness region of the verification technique.
However, the following slight relaxation of formula (\ref{p3a}) can be proved to hold for the system:
%
%
%

\begin{equation}\label{p3b}
  \boxMTL{(0,T_3+\delta]}{\tait{data\_retrieved}} \Rightarrow \diamondMTL{(0, T]}{\tait{try}}
\end{equation}

%

\begin{table}[!htb]
\label{tab:exp_res}
\begin{scriptsize}
\begin{center}
  \begin{tabular}{|c | c c c c c | c c c c c|}    
    \hline
    \textsc{Pr} & \textsc{$T_1$} & \textsc{$T_2$} & \textsc{$T_3$} & \textsc{$T$} & \textsc{k} & \textsc{Pre} (min.) & \textsc{CNF} (hrs.) & \textsc{SAT} (hrs.) & \textsc{$\naturals$-valid} & \textsc{\# Cl}$ \cdot 10^6$ \\
    \hline

    \ref{p2a}: $\phi^+$ &  3 & 6 & 18 & 30 & 90 & 1.9877 & 1.854 & 2.2322 & $\perp$ & 12.4148\\
    \ref{p2a}: $\phi^-$ &  3 & 6 & 18 & 30 & 90 & 3.0743 & 6.2533 & 5.3518 & $\perp$ & 21.306\\
    \ref{p2a}: $\phi^+$ &  3 & 9 & 36 & 30 & 90 & 2.425 & 2.5699 & 2.5368 & $\perp$ & 12.7411\\
    \ref{p2a}: $\phi^-$ &  3 & 9 & 36 & 30 & 90 & 3.3372 & 6.2202 & 5.0851 & $\perp$ & 21.6323\\
    \ref{p2a}: $\phi^+$ &  3 & 12 & 48 & 30 & 120 & 3.2059 & 3.6904 & 6.8226 & $\perp$ & 17.2833\\
    \ref{p2a}: $\phi^-$ &  3 & 12 & 48 & 30 & 120 & 4.5452 & 10.439 & 9.2688 & $\perp$ & 29.117\\

    \hline

    \ref{p2b}: $\phi^+$ &  3 & 6 & 18 & 30 & 90 & 2.1074 & 1.9171 & 0.8101 & $\top$ & 12.8512\\
    \ref{p2b}: $\phi^-$ &  3 & 6 & 18 & 30 & 90 & 3.1059 & 5.7381 & 3.017 & $\top$ & 21.7514\\
    \ref{p2b}: $\phi^+$ &  3 & 9 & 36 & 30 & 90 & 2.6346 & 2.7726 & 0.9741 & $\top$ & 13.1775\\
    \ref{p2b}: $\phi^-$ &  3 & 9 & 36 & 30 & 90 & 3.5125 & 6.4452 & 3.5557 & $\top$ & 22.0778\\
    \ref{p2b}: $\phi^+$ &  3 & 12 & 48 & 30 & 120 & 3.6731 & 4.379 & 2.0955 & $\top$ & 17.8641\\
    \ref{p2b}: $\phi^-$ &  3 & 12 & 48 & 30 & 120 & 5.1492 & 11.0093 & 5.0007 & $\top$ & 29.7098\\

    \hline

    \ref{p1a}: $\phi^+$ &  3 &  6 &  18 &  30 & 90 & 1.8887 & 1.7376 & 3.1524 & $\perp$  & 12.0598\\
    \ref{p1a}: $\phi^-$ &  3 &  6 &  18 &  30 & 90 & 2.9094 & 6.0154 & 3.427 & $\top$ & 20.931\\
    \ref{p1a}: $\phi^+$ &  3 &  9 &  36 &  30 & 90 & 2.2002 & 2.3232 & 2.4845 & $\perp$  & 12.3862\\
    \ref{p1a}: $\phi^-$ &  3 &  9 &  36 &  30 & 90 & 3.1067 & 5.8341 & 4.6997 & $\top$ & 21.2573\\
    \ref{p1a}: $\phi^+$ &  3 &  12 &  48 &  30 & 120 & 3.4446 & 4.1686 & 8.8680 & $\perp$ & 16.8108\\
    \ref{p1a}: $\phi^-$ &  3 &  12 &  48 &  30 & 120 & 4.1621 & 9.9533 & 13.1718 & $\top$ & 28.6179\\
    
    \hline

    \ref{p1c}: $\phi^+$ &  3 & 6 & 18 & 30 & 90 & 2.0715 & 1.6828 & 1.2976 & $\top$ & 12.1584\\
    \ref{p1c}: $\phi^-$ &  3 & 6 & 18 & 30 & 90 & 3.0536 & 5.3665 & 3.9414 & $\top$ & 21.0296\\
    \ref{p1c}: $\phi^+$ &  3 & 9 & 36 & 30 & 90 & 2.8152 & 2.2134 & 1.7645 & $\top$ & 12.4848\\
    \ref{p1c}: $\phi^-$ &  3 & 9 & 36 & 30 & 90 & 3.7314 & 6.1665 & 3.6802 & $\top$ & 21.3559\\
    \ref{p1c}: $\phi^+$ &  3 & 12 & 48 & 30 & 120 & 3.9268 & 4.5246 & 9.3435 & $\perp$ & 16.9421\\
    \ref{p1c}: $\phi^-$ &  3 & 12 & 48 & 30 & 120 & 4.8244 & 9.7484 & 14.8257 & $\top$ & 28.7491\\

    \hline

    \ref{p1a}': $\phi^+$ &  3 & 6 & 18 & 30 & 90 & 2.2399 & 2.3971 & 4.0335 & $\top$ & 12.8097\\
    \ref{p1a}': $\phi^-$ &  3 & 6 & 18 & 30 & 90 & 3.3884 & 5.5905 & 4.5752 & $\top$ & 21.6645\\
    \ref{p1a}': $\phi^+$ &  3 & 9 & 36 & 30 & 90 & 2.4788 & 2.2978 & 4.8259 & $\top$ & 13.136\\
    \ref{p1a}': $\phi^-$ &  3 & 9 & 36 & 30 & 90 & 3.8369 & 7.3132 & 0.0036 & $\top$ & 21.9909\\
    \ref{p1a}': $\phi^+$ &  3 & 12 & 48 & 30 & 120 & 4.7220 & 5.0607 & 13.3136 & $\perp$ & 17.8088\\
    \ref{p1a}': $\phi^-$ &  3 & 12 & 48 & 30 & 120 & 4.8557 & 9.7088 & 8.4951 & $\top$ & 29.5942\\

    \hline

    \ref{p3}: $\phi^+$ &  3 & 6 & 12 & 30 & 75 & 1.5108 & 1.0502 & 0.4716 & $\perp$ & 9.91056\\
    \ref{p3}: $\phi^-$ &  3 & 6 & 12 & 30 & 75 & 2.1418 & 3.1694 & 1.4723 & $\perp$ & 17.3177\\
    \ref{p3}: $\phi^+$ &  3 & 3 & 15 & 30 & 75 & 1.5199 & 1.0564 & 0.4703  & $\perp$ & 9.87584\\
    \ref{p3}: $\phi^-$ &  3 &  3 &  15 &  30 & 75 & 2.1586 & 3.1764 & 1.4473 & $\perp$ & 17.2837\\
    \ref{p3}: $\phi^+$ &  3 & 6 & 18 & 30 & 75 & 1.5458 & 1.0706 & 0.5673 & $\perp$ & 9.97865\\
    \ref{p3}: $\phi^-$ &  3 &  6 &  18 &  30 & 75 & 2.1978 & 3.2174 & 1.4323 & $\perp$ & 17.3858\\

    \hline

    \ref{p3a}: $\phi^+$ &  3 &  6 &  12 &  30 & 75 & 1.6018 & 1.1108 & 0.8844  & $\perp$  & 9.97312\\
    \ref{p3a}: $\phi^-$ &  3 &  6 &  12 &  30 & 75 & 2.2909 & 3.3455 & 2.1095 & $\top$  & 17.3841\\
    \ref{p3a}: $\phi^+$ &  3 &  3 &  15 &  30 & 75 & 1.6734 & 1.1945 & 0.6418 & $\perp$ & 9.95542\\
    \ref{p3a}: $\phi^-$ &  3 &  3 &  15 &  30 & 75 & 2.1638 & 3.2626 & 1.5792 & $\top$  & 17.3671\\
    \ref{p3a}: $\phi^+$ &  3 &  6 &  18 &  30 & 75 & 1.7031 & 1.2210 & 0.9653 & $\top$ & 10.0752\\
    \ref{p3a}: $\phi^-$ &  3 & 6 & 18 & 30 & 75 & 2.48 & 3.3642 & 1.1761  & $\top$ & 17.4862\\

    \hline

    \ref{p3b}: $\phi^+$ &  3 &  6 &  12 &  30 & 75 & 1.578 & 1.0879 & 1.2972 & $\top$ & 9.97879\\
    \ref{p3b}: $\phi^-$ &  3 & 6 & 12 & 30 & 75 & 2.3035 & 3.2128 & 1.6002 & $\top$ & 17.3898\\
    \ref{p3b}: $\phi^+$ &  3 &  3 &  15 &  30 & 75 & 1.6465 & 1.0986 & 0.7740 & $\perp$  & 9.96109\\
    \ref{p3b}: $\phi^-$ &  3 &  3 &  15 &  30 & 75 & 2.1604 & 3.1919 & 1.1408 & $\top$ & 17.3727\\
    \ref{p3b}: $\phi^+$ &  3 &  6 &  18 &  30 & 75 & 1.6220 & 1.1249 & 0.8240 & $\top$ & 10.0809\\
    \ref{p3b}: $\phi^-$ &  3 & 6 & 18 & 30 & 75 & 2.2892 & 3.2682 & 1.1178 & $\top$ & 17.4919\\

%
    
    \hline

  \end{tabular}
\caption{Checking properties of the data monitoring system.}
\end{center}
\end{scriptsize}
\end{table}


%% file: Conclusion.tex
\section{Discussion and Conclusion}\label{sec:conclusion}

In this paper we presented a technique to formally model and verify systems using different paradigms for different system parts.
The technique hinges on MTL axiomatizations of the different modeling notations, which provide a common formal ground for the various modeling languages, on which fully-automated verification techniques are built.
We provided an MTL axiomatization of a subset of TPN, a typical asynchronous operational formalism, and showed how models could be built by formally combining together TPN and TA (a classic synchronous operational notation, for which an axiomatization has been provided in \cite{FPR08-ICFEM08}).
In addition, we showed how the approach allows users to integrate in the same model parts described through a continuous notion of time, and parts described through a discrete notion of time.

Practical verification of systems modeled through the multi-paradigm approach is possible through the \zot{} bounded satisfiability checker, for which plugins supporting the various axiomatized notations have been built.

The technique has been validated on a non trivial example of data monitoring system.
The experimental results show the feasibility of the approach, through which we have been able to investigate the validity (or, in some cases, the non validity) of some properties of the system.
As described in Section \ref{sec:casestudy}, the verification phase has provided useful insights on the mechanisms and on the timing features of the modeled system, which led us to re-evaluate some of our initial beliefs on the system properties.

It is clear from our experiments that, unsurprisingly, the technique suffers
from two main drawbacks: the incompleteness of the verification approach by
discretization evidenced in \cite{FPR08-FM08}, which prevented us, in some
cases, to get conclusive answers on some analyzed properties; and the
computational complexity of our method, which is based on the
direct translation of TPN and TA into MTL formulas, approximated into
discrete ones, and then encoded into SAT. This makes proofs considerably lengthier as the size of the domains, and especially of the temporal one, increases, as evidenced by Table \ref{tab:exp_res}.
Nevertheless, we maintain that the results we obtained are promising, and show the applicability of the technique on non trivial systems.
This claim is supported on the one hand by the sophistication of the properties we have been able to prove (or disprove): it is inevitable that verification over continuous real-time has a high computational cost.
On the other hand, while incompleteness is a hurdle to the full applicability of the technique, in practice it can be mitigated quite well, usually by slightly relaxing the real-time timing requirements under verification in a way that does not usually alter the gist of what is being verified.

In our future research on this topic we plan to address the two main drawbacks evidenced above.
First, we will work on extending the verification technique to expand its range of applicability and reduce its region of incompleteness.
Also, we will study more efficient implementations for the \zot{} plugins
through which the various modeling notations are added to the framework: we
believe that more direct (therefore more compact, both in the literals and
clause numbers) encodings into SAT of the TPN and TA axiomatizations
should significantly improve the efficiency of the tool.

In particular, we have not yet tackled the problem of optimizing the encodings of the TPN and TA axiomatizations into the SAT problem.
We expect that significant improvements on the duration of the proofs can be gained through optimized encodings that reduce, on the one hand, the time needed to put formulas in the conjunctive normal form that is required as input by SAT solvers, and, on the other hand, the number of literals required to represent TPN and TA as SAT problems.


%% file: IntModVerRTMP.bbl
\begin{thebibliography}{10}

\bibitem{AD94}
R.~Alur and D.~L. Dill.
\newblock A theory of timed automata.
\newblock {\em Theor. Comp. Sci.}, 126(2):183--235, 1994.

\bibitem{AFH96}
R.~Alur, T.~Feder, and T.~A. Henzinger.
\newblock The benefits of relaxing punctuality.
\newblock {\em Journal of the ACM}, 43(1):116--146, 1996.

\bibitem{AH92b}
R.~Alur and T.~A. Henzinger.
\newblock Logics and models of real time: A survey.
\newblock In {\em Proc.~of Real-Time: Theory in Practice}, volume 600 of {\em
  LNCS}, pages 74--106, 1992.

\bibitem{AH93}
R.~Alur and T.~A. Henzinger.
\newblock Real-time logics: Complexity and expressiveness.
\newblock {\em Information and Computation}, 104(1):35--77, 1993.

\bibitem{CR06}
F.~Cassez and O.~H. Roux.
\newblock Structural translation from time {P}etri nets to timed automata.
\newblock {\em Journal of Systems and Software}, 79(10):1456--1468, 2006.

\bibitem{CMS99}
A.~Cerone and A.~Maggiolo-Schettini.
\newblock Time-based expressivity of time {P}etri nets for system
  specification.
\newblock {\em Theor. Comp. Sci.}, 216(1--2):1--53, 1999.

\bibitem{CGP00}
E.~M. Clarke, O.~Grumberg, and D.~A. Peled.
\newblock {\em Model Checking}.
\newblock MIT Press, 2000.

\bibitem{FMM94}
M.~Felder, D.~Mandrioli, and A.~Morzenti.
\newblock Proving properties of real-time systems through logical
  specifications and {P}etri net models.
\newblock {\em {IEEE} Trans. on Soft. Eng.}, 20(2):127--141, 1994.

\bibitem{FMMR07-TR2007-22}
C.~A. Furia, D.~Mandrioli, A.~Morzenti, and M.~Rossi.
\newblock Modeling time in computing.
\newblock Technical Report 2007.22, DEI, Politecnico di Milano, January 2007.

\bibitem{FMMR09}
C.~A. Furia, D.~Mandrioli, A.~Morzenti, and M.~Rossi.
\newblock Modeling time in computing: a taxonomy and a comparative survey.
\newblock {\em ACM Computing Surveys}, to appear.
\newblock Also available as \texttt{http://arxiv.org/abs/0807.4132}.

\bibitem{FPR08-FM08}
C.~A. Furia, M.~Pradella, and M.~Rossi.
\newblock Automated verification of dense-time {MTL} specifications via
  discrete-time approximation.
\newblock In {\em Proc.~of FM'08}, volume 5014 of {\em LNCS}, pages 132--147,
  2008.

\bibitem{FPR08-ICFEM08}
C.~A. Furia, M.~Pradella, and M.~Rossi.
\newblock Practical automated partial verification of multi-paradigm real-time
  models.
\newblock In {\em Proc.~of ICFEM'08}, volume 5256/-1 of {\em LNCS}, pages
  298--317, 2008.

\bibitem{FR06}
C.~A. Furia and M.~Rossi.
\newblock Integrating discrete- and continuous-time metric temporal logics
  through sampling.
\newblock In {\em Proc.~of FORMATS'06}, volume 4202 of {\em LNCS}, pages
  215--229, 2006.

\bibitem{HM96}
C.~Heitmeier and D.~Mandrioli, editors.
\newblock {\em Formal Methods for Real-Time Computing}.
\newblock John Wiley \& Sons, 1996.

\bibitem{Koy90}
R.~Koymans.
\newblock Specifying real-time properties with metric temporal logic.
\newblock {\em Real-Time Systems}, 2(4):255--299, 1990.

\bibitem{zot}
M.~Pradella.
\newblock \zot{}.
\newblock \texttt{http://home.dei.polimi.it/ pradella}, March 2007.

\bibitem{PMS07}
M.~Pradella, A.~Morzenti, and P.~{San Pietro}.
\newblock The symmetry of the past and of the future: bi-infinite time in the
  verification of temporal properties.
\newblock In {\em Proc. of ESEC/FSE 2007}, 2007.

\bibitem{UML}
{OMG Unified Modeling Language (OMG UML) Superstructure}, v2.2.
\newblock Technical Report formal/2009-02-02, Object Management Group, 2009.

\bibitem{vdB94}
M.~{von der Beeck}.
\newblock A comparison of statecharts variants.
\newblock In {\em Proc. of FTRTFT}, volume 863 of {\em LNCS}, pages 128--148,
  1994.

\end{thebibliography}
